\documentclass[8pt,a4paper]{article}
\usepackage[utf8]{inputenc}
\usepackage[english]{babel}
\usepackage[T1]{fontenc}
\usepackage{geometry}
\usepackage{setspace}
\usepackage{tocloft}
\usepackage{tikz}
\usetikzlibrary{calc,arrows}
\usepackage{multicol}
\usepackage{float}
\usepackage{lipsum} 
\usepackage{amsmath}
\usepackage{mathptmx}
\usepackage{caption}
\usepackage{etoolbox}
\usepackage{booktabs}
\usepackage{adjustbox}
\usepackage{multicol}
\usepackage{algorithm}
\usepackage{algpseudocode}

\usepackage{fancyhdr}
\definecolor{WNEcolor}{HTML}{C00000}

\usepackage{afterpage}

\usepackage{authblk}

\begin{document}

\title{\textbf{\small Supervised Autoencoder MLP for Financial Time Series Forecasting}}

\vspace{8.0cm}

\author[1]{\small Bartosz Bieganowski}

\author[2]{Robert Ślepaczuk}

\affil[1]{\small University of Warsaw, Faculty of Economic Sciences, Quantitative Finance Research Group, Ul. Długa 44/50, 00-241 Warsaw, Poland. ORCID: https://orcid.org/0009-0003-8671-8266\vspace{0.5cm}}

\affil[2]{\small University of Warsaw, Faculty of Economic Sciences, Department of Quantitative Finance and Machine Learning, Quantitative Finance Research Group, Ul. Długa 44/50, 00-241 Warsaw, Poland. ORCID: https://orcid.org/0000-0001-5527-2014, corresponding author: rslepaczuk@wne.uw.edu.pl}

\date{}

\maketitle

\begin{abstract}
This paper investigates the enhancement of financial time series forecasting with the use of neural networks through supervised autoencoders, aiming to improve investment strategy performance. It specifically examines the impact of noise augmentation and triple barrier labeling on risk-adjusted returns, using the Sharpe and Information Ratios. The study focuses on the S\&P 500 index, EUR/USD, and BTC/USD as the traded assets from January 1, 2010, to April 30, 2022. Findings indicate that supervised autoencoders, with balanced noise augmentation and bottleneck size, significantly boost strategy effectiveness. However, excessive noise and large bottleneck sizes can impair performance, highlighting the importance of precise parameter tuning. This paper also presents a derivation of a novel optimization metric that can be used with triple barrier labeling. The results of this study have substantial policy implications, suggesting that financial institutions and regulators could leverage techniques presented to enhance market stability and investor protection, while also encouraging more informed and strategic investment approaches in various financial sectors. 

\end{abstract}

\vspace{10.0cm}
\bigskip
\noindent \textit{Note:} This research did not receive any specific grant from funding agencies in the public, commercial, or not-for-profit sectors.

\setlength{\parindent}{0pt} 

\singlespacing
\setlength{\parindent}{1cm}
\setstretch{1}
\newpage
\section*{Introduction}
\addcontentsline{toc}{section}{Introduction}

This research focuses on the development of an Algorithmic Investment Strategy (AIS) that leverages Supervised Autoencoder - Multi-Layer Perceptron (SAE-MLP) networks. This strategy aims to utilize high-frequency price data, a significant shift from traditional methods that primarily rely on daily closing prices. The study is structured to test and answer three primary research questions:

\begin{itemize}
    \item RQ1. Does data augmentation and denoising via autoencoders improve the performance of a strategy?
    \item RQ2. Does triple barrier labeling improve classifier performance over simple direction classification?
    \item RQ3. Does hyperparameter tuning help achieve better performance of the
investment strategy?
\end{itemize}

The financial instruments under study include the S\&P 500 Index (SPX), the EUR/USD currency pair, and Bitcoin (BTC). Our dataset spans from January 1, 2016, to April 31, 2022, with the latter period used as out-of-sample data for testing. It is important to note the difference in trading hours for these assets, with the S\&P 500 index having restricted hours and Bitcoin being traded 24/7.

To address our research questions, we employ empirical methods, developing trading strategies based on various SAE-MLP model architectures. These models are tested using data collected at one-minute intervals. Our approach involves using the SAE-MLP model for price prediction, employing a walk-forward method. This process begins with hyperparameters sourced from existing literature, followed by fine-tuning through hyperparameter optimization. We anticipate that models with algorithm-selected hyperparameters will show superior performance.

Our methodology also includes transforming the return estimation problem from a regression model to a classification one, focusing on predicting price direction rather than the exact price. This shift is expected to be more effective for generating investment signals due to the more relevant loss function in this context. We subsequently test triple barrier labeling, a novel technique of estimating maximum and minimum prices from a given period instead of fixed-horizon close prices. Additionally, we conduct a sensitivity analysis to examine the robustness of our findings under different assumptions.

The structure of this paper is as follows: Section 1 provides a literature review, emphasizing neural network applications in stock and cryptocurrency price predictions. Section 2 details the data and financial instruments used. Section 3 describes the methodology, including the SAE-MLP model, walk-forward approach, equity line construction, ensemble model, and performance metrics. It also elaborates on the research approaches. Section 4 presents empirical results, comparing them with each other through equity line charts and performance metrics tables. Section 5 conducts a sensitivity analysis of the most promising approach, focusing on variations in hyperparameters. The paper concludes with Section 6, summarizing the research findings.

\section{Literature Review}
The exploration of financial market efficiency and the predictability of stock price movements has long captivated academic interest. This has led to extensive research integrating machine learning techniques into financial analysis. Central to this discussion is the Efficient Market Hypothesis (EMH), which posits that stock prices reflect all available information, rendering them fundamentally unpredictable. This hypothesis is categorized into three forms: weak, semi-strong, and strong. The weak form challenges the basis of technical analysis by suggesting that stock prices follow a random path. The semi-strong form argues that stock prices instantaneously incorporate all public information, thereby diminishing the effectiveness of fundamental analysis. The strong form extends this argument to include all information, public and private, in the determination of stock prices (Malkiel, 1973).

Fama (1970) acknowledged the presence of statistically significant patterns in daily returns that could potentially affect profitable trading strategies, yet he also noted that these patterns might not contradict the EMH. Malkiel (2005) further argued that active equity management often fails to outperform passive investment strategies, a viewpoint that aligns with the EMH. However, behavioral finance theorists like Barberis and Thaler (2002) have countered by suggesting that market inefficiencies can cause stock prices to deviate from their true values. This perspective challenges the notion that the inability of portfolio managers to consistently beat market indices is a definitive indicator of market efficiency. Contrary to the EMH, a significant contingent of technical analysts, or technicians, maintain that historical price data can be used to forecast future stock prices (Lo and Hasanhodzic, 2010). Additionally, the existence of stock market anomalies and serial correlations in economic factors presents further challenges to the EMH (Abu-Mostafa and Atiya, 1996). Another prominent model in time series prediction is the Box-Jenkins Model, or ARIMA (autoregressive integrated moving average), formulated by George Box and Gwilym Jenkins. This model combines autoregression, differencing, and moving averages to predict stock prices. It has been widely applied in financial forecasting (Box and Jenkins, 1976).

Given the recent advancements in powerful computers and efficient machine learning algorithms, it is not surprising that many machine learning (ML) techniques have been researched for forecasting the direction of financial instruments’ prices. Examples of supervised learning approaches that can be trained to forecast asset prices and trends based on previous data and provide insightful historical price analysis include Support Vector Machine (SVM), Decision Trees, and ANNs (Zenkova and Ślepaczuk, 2019). More papers trying to tackle this issue are described below.

Ariyo et al. (2014) applied the ARIMA model to data from the New York and Nigeria Stock Exchanges, demonstrating its capability in short-term price prediction and its comparability to other techniques like ANNs. Azari (2018) also tested the ARIMA model's effectiveness in predicting Bitcoin values over three years. The model showed proficiency in short-term forecasts, particularly in stable periods, but struggled with sudden price fluctuations and long-term predictions, especially in volatile periods like the end of 2017. This limitation highlights the challenges in predicting prices over extended periods or in highly volatile markets. In a study conducted by Siami et al. (2018), the effectiveness of deep learning-based time series forecasting algorithms, including LSTM, was compared to that of more established algorithms like ARIMA. The results showed that LSTM and other deep learning-based techniques outperformed ARIMA with an average error rate reduction of 84 to 87 percent, demonstrating the superiority of LSTM. The study also found that the trained forecast model behaved randomly and the number of training cycles (epochs) had no impact on its performance. In another study by Grudniewicz and Ślepaczuk (2021), various Machine Learning algorithms were applied to technical analysis indicators for the WIG20, DAX, S\&P 500, and a few selected CEE indices. The study concluded that quantitative techniques outperformed passive strategies in terms of risk-adjusted returns, with the Bayesian Generalized Linear Model and Naive Bayes being the top models for the investigated indices. Di Persio and Honchar (2017) investigated the performance of three recurrent neural network (RNN) models, including a basic RNN, LSTM, and Gated Recurrent Unit (GRU), using the price of Google stock as input. The study also provided insights into the hidden dynamics of RNN. Their results showed that the LSTM outperformed the other models with a 72 percent accuracy on a five-day horizon. Kijewski and Ślepaczuk (2020) utilized traditional techniques and a recurrent neural network model (LSTM) to implement buy/sell signals for algorithmic investment strategies. Their study evaluated the effectiveness of investment algorithms on the S\&P 500 index time series, spanning 20 years of data from 2000 to 2020. They employed a rolling training-testing window for dynamic parameter optimization throughout the backtesting process. The combination of signals from several methods doubled the returns on the same level of risk of the Buy \& Hold strategy benchmark. The study further found that the LSTM model was significantly less resistant to changes in parameters than conventional techniques, as demonstrated through a thorough sensitivity study.

Many studies have explored the use of ensemble or hybrid techniques with LSTM for improved performance. Hossain et al. (2018) developed a deep learning-based hybrid model consisting of LSTM and GRU architectures to predict S\&P 500 prices using a dataset spanning 66 years. In this approach, the input data is passed to the LSTM network to generate a first-level prediction, and the output of the LSTM layer is then passed to the GRU layer to provide the final prediction. The proposed model outperformed earlier neural network techniques with a Mean Squared Error (MSE) of 0.00098 in prediction.

However, many current studies do not consider non-stationarity, which is a significant challenge in financial time series forecasting. Shah et al. (2018) demonstrated the potential of LSTM-RNN-based models in predicting non-stationary data. The study shows that the LSTM model provides excellent results for daily forecasts and satisfactory outcomes for predictions made seven days in advance using only daily price as a feature. The authors used a larger training dataset spanning 20 years, including market ups and downs, to train the model. The study suggests that LSTM RNN has the potential for identifying underlying trends and producing longer-term forecasts with the addition of more features, particularly for volatile stock datasets.

Baranochnikov and Ślepaczuk (2022) propose a walk-forward procedure for cross-validation of machine learning models in financial time series forecasting. They apply this method to test algorithms on four financial assets (Bitcoin, Tesla, Brent Oil, and Gold) and conclude that LSTM outperforms GRU in most cases. De Prado (2016) presented a method called combinatorial purged cross-validation (CV) for backtesting time series data that addresses some of the issues with traditional k-fold CV methods, such as that finance data cannot be expected to be drawn from an independent and identically distributed (IID) process, causing k-fold CV to fail. Additionally, the testing set is often used multiple times during model development, leading to selection bias. 
In their study, Kamalov et al. (2021) analyze the effectiveness of stock price and return as input features for directional forecasting models, using data from ten high-cap US corporations over ten years. They employ four categorization techniques to construct forecasting models and observe that stock price outperforms return as an independent input feature for predicting the direction of price movement. However, when technical indicators are added to the input feature set, the performance difference between the two input features diminishes. The authors conclude that stock price is a more potent input feature than return value in predicting the direction of price movement.

Le et al. (2018) propose a novel approach for improving the generalization performance of neural networks through the use of a supervised auto-encoder. The authors note that regularizing hidden layers in neural networks is not a straightforward task, and existing layer-wise suggestions lack theoretical guarantees for improved performance. In their work, they analyze the supervised auto-encoder model, which predicts both inputs (reconstruction error) and targets jointly. The authors provide a novel generalization results for linear auto-encoders, proving uniform stability based on the inclusion of the reconstruction error. Their approach is shown to outperform simplistic regularization methods such as norms, as well as more advanced regularization techniques such as the use of auxiliary tasks. 

The use of Machine Learning techniques for predicting highly nonlinear and noisy data on digital blockchain platforms has become increasingly common due to the notable increase in cryptocurrency trading. In their study, Suhwan et al. (2019) examined several cutting-edge deep learning techniques for predicting Bitcoin prices, including deep neural networks (DNN), long short-term memory (LSTM) models, convolutional neural networks, deep residual networks, and their combinations. The experimental results showed that DNN-based models outperformed the other models in predicting the direction of price movement, while LSTM-based models outperformed the other models in price prediction. Furthermore, the evaluation of profitability indicated that classification models were superior to regression models in algorithmic trading.

Lahmiri and Bekiros (2020) examine the use of machine learning (ML) models in high-frequency trading of Bitcoin. They explore three different types of models, including algorithmic models such as regression trees, statistical ML approaches like support vector regressions (SVR), and artificial neural network (ANN) topologies such as feedforward (FFNN) or Bayesian regularization (BRNN). The study's results suggest that ANN models outperform other models in noisy signal environments. The authors argue that the increasing need for understanding and forecasting variables across shorter time horizons, coupled with technological advancements and big data processing power, has led to the current developments in high-frequency data estimation. Michańków et al. (2022) projected the future values of the Bitcoin and S\&P 500 index, utilizing data from 2013 to the end of 2020 and across daily, hourly, and 15-minute frequency intervals. They formulated a unique loss function, which amplifies the predictive capabilities of the LSTM model for algorithmic investment strategies. The researchers determined that the primary elements dictating the efficacy of LSTM in algorithmic investment tactics include the approach adopted for hyperparameter tuning, the architecture of the model, and the estimation process.

In the spot and futures markets for the S\&P 500, Schulmeister (2009) examines how technical trading strategies can use momentum and reversal effects. Based on daily statistics, 2580 technical models’ profitability has continuously decreased since 1960 and has been negative since the early 1990s. The same models, however, yield an average gross return of 7.2\% each year between 1983 and 2007 when based on 30-minute data. This outcome may be the consequence of recent improvements in stock market efficiency or a change in stock price trends from 30-minute prices to higher frequency pricing, a claim later supported by Kryńska and Ślepaczuk (2022).

Dudek et al. (2024) perform a comparative analysis of statistical and machine learning methods for forecasting cryptocurrency volatility, examining models like GARCH, LASSO, and LSTM across Bitcoin, Ethereum, Litecoin, and Monero. They find that no single model is universally best, but SVR and FNM show consistent performance across different cryptocurrencies and time frames. Ozbayoglu et al. (2020) provide a comprehensive survey of deep learning applications in finance, categorizing works by financial subfields and analyzing them based on their deep learning models. They highlight the significant potential of deep learning in finance and suggest areas for future research. Sezer et al. (2020) offer a systematic literature review on deep learning for financial time series forecasting, categorizing studies by their forecasting areas and deep learning models. They emphasize the superior performance of deep learning over traditional models and identify research gaps and opportunities. Tkáč and Verner (2016) review the use of artificial neural networks in business applications over two decades, focusing on financial distress forecasting, stock price prediction, and decision support. They note a preference for multilayer feedforward networks and suggest ample room for further research to enhance neural network models.

Hu et al. (2015) conducted a literature review on evolutionary computation for rule discovery in stock algorithmic trading, noting a bias toward genetic algorithms and technical analysis. They identify research gaps in fundamental analysis, transaction cost consideration, and the combination of trading rule discovery with portfolio selection. Fischer and Krauss (2018) demonstrate the effectiveness of LSTM networks in predicting financial market movements, outperforming traditional models with substantial profits before 2010. They note that LSTM networks profit from stocks with high volatility and short-term reversal characteristics, and show low exposure to systematic risk. Krauss et al. (2017) evaluate deep neural networks, gradient-boosted trees, and random forests for statistical arbitrage on the S\&P 500, finding that an ensemble of these methods yields significant out-of-sample returns. This poses a challenge to the semi-strong form of market efficiency, despite a recent decline in profits. 
Sezer and Ozbayoglu (2018) propose an algorithmic trading model using deep convolutional neural networks, converting financial time series into 2-D images through technical indicators. Their model outperforms traditional trading systems and the buy-and-hold strategy, showcasing the potential of deep learning in financial forecasting.

To summarize, many studies indicate that machine learning models outperform other techniques, particularly statistical approaches (e.g., ARIMA models), especially for nonstationary data. It has been proposed that models perform better when the input value is asset price rather than return. It has further been suggested that intra-day data-driven strategies should outperform inter-day data-driven strategies. In our study, we aim to
enhance the research by comparing different types of neural network models to develop successful trading strategies using prices and other features as input data. 

\section{Data}

\subsection{Traded Assets}

The research presented in this paper uses three time series of data for trading and six more as features. The first traded time series refers to the Standard and Poor's 500 (S\&P500) index. Minute-frequency open-high-low-close (OHCL) data is used, which was obtained commercially from FirstRate Data. The data ranges from January 2010 to April 2022. The data before January 1st, 2020 is used exclusively for training strategies. Using long-range training allows for testing the investment algorithms through a variety of market regimes, with different combinations of trend and volatility. It includes periods of high volatility and downside shocks, such as the financial crisis of 2008, and the consecutive uptrend of 2009. The same point can be made for test data - the COVID-related downturn in March of 2020, the rebound in the following months, as well as steady bear market of 2022 make the testing environment of trading algorithms quite robust. Figure 1 presents fluctuations of SP500 in the period of our research.\\

The second traded asset is Bitcoin (Figure 2), the cryptocurrency that was released in 2009, which only years later gained significant popularity as a tradeable asset. The popularity is linked to the rapid growth of the price of Bitcoin in its lifetime, caused mainly by expectations for Bitcoin to be a successor of fiat money. The data was also obtained from FirstRate Data in OHLC format. The testing data for Bitcoin covers two of the major "peaks" in its price, which provides a way to test the algorithm's robustness in extreme conditions.\\

\begin{figure}[ht]
  \centering
  \begin{minipage}{0.49\textwidth}
    \centering
    \includegraphics[width=1\textwidth]{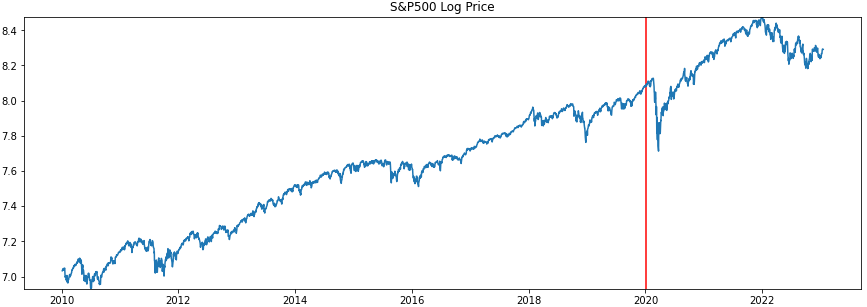} 
    \caption{Log price of SP500 with train/test cutoff.}
    \caption*{\footnotesize Source: FirstRateData, S\&P 500 index in the period from January 1, 2010 to April 30, 2022; Accessed May 2022.}
    \label{fig:spx}
  \end{minipage}
  \hfill 
  \begin{minipage}{0.49\textwidth}
    \centering
    \includegraphics[width=1\textwidth]{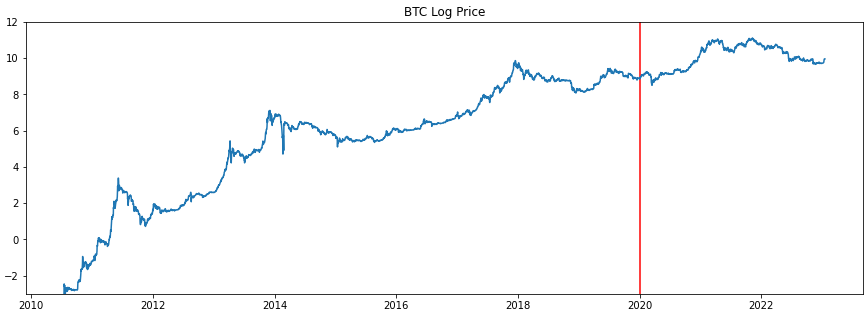} 
    \caption{Log price of BTCUSD with train/test cutoff.}
    \caption*{\footnotesize Source: FirstRateData, BTC/USD pair in the period from January 1, 2010, to April 30, 2022; Accessed May 2022.}
    \label{fig:btc}
  \end{minipage}
\end{figure}

The third traded asset is the EUR/USD pair (Figure 3), which shares some characteristics with both Bitcoin (tradeable 24/7) and SP500 (being one of the most liquid assets in the world). Although the test period represents mostly a downward trend, we view it as a valuable balancing asset to test as compared to the mostly growing S\&P 500 index and BTC/USD.\\

\begin{figure}
  \centering
  \caption{Price of EURUSD with train/test cutoff.}
      \includegraphics[width=0.5\columnwidth]{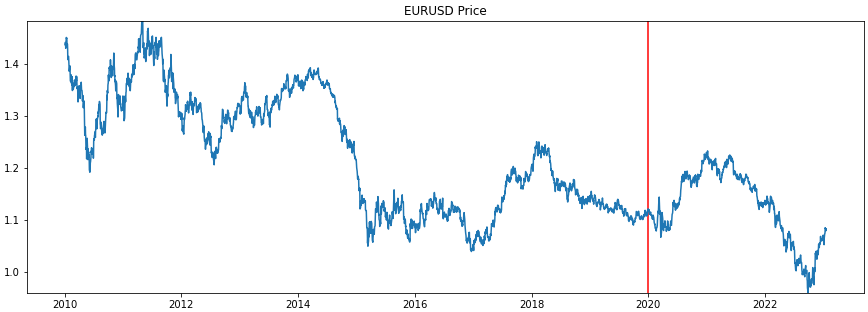}
  \caption*{\footnotesize Source: FirstRateData, EUR/USD pair in the period from January 1, 2010, to April 30, 2022; Accessed May 2022.}
\end{figure}

Each of the three assets varies in terms of fundamental characteristics, such as asset class and the number of participants in the market, but also in terms of statistical properties of their returns. The descriptive statistics obtained for the traded financial instruments at selected frequencies are shown in Table 1. We can notice that the average return across all frequencies and for all three instruments is positive. In terms of volatility, Bitcoin is clearly the most volatile asset, followed by S\&P 500 and EURUSD in the last place.

\begin{table}[tb]
\centering
\caption{Descriptive statistics for daily returns of the three traded assets, daily data (2010-01-01 to 2022-04-30).}
\begin{tabular}{lrrr}
\toprule
 & EURUSD & BTC & SPX \\
\midrule
count & 3147 & 4305 & 3147 \\
mean  & -0.00352\% & 0.52245\% & 0.0481\% \\
std   & 0.53612\% & 7.78643\% & 1.1135\% \\
min   & -2.90496\% & -57.20567\% & -11.9840\% \\
25\%   & -0.31678\% & -1.35672\% & -0.3770\% \\
50\%   & 0.00739\% & 0.16627\% & 0.0612\% \\
75\%   & 0.30343\% & 2.13115\% & 0.5626\% \\ 
max   & 3.00608\% & 336.83900\% & 9.3827\% \\
\bottomrule
\end{tabular}
\vspace{0.5cm}
\caption*{\footnotesize Source: Own Elaboration based on FirstRateData, Accessed May 2022}
\end{table}

\subsection{Feature Time Series}

When it comes to the time series used as features in this paper, the first one is the seasonally adjusted Initial Claims (ICSA), retrieved from the Federal Reserve Bank of St. Louis, FRED. An initial claim is a claim filed by an unemployed individual after a separation from an employer. The claim requests a determination of basic eligibility for the Unemployment Insurance program. This data is released every Thursday by the FED and reports the situation of unemployment on the last Saturday.

West Texas Intermediate (WTI) Light Sweet Crude Oil futures are currently the world’s most liquid oil contract. The inclusion of this asset as a feature is supported by the correlation between some of the traded assets and oil. A rise in oil prices can indicate strong economic growth and increased demand for energy, which can influence each of the traded assets. Conversely, a fall in oil prices can indicate a slowdown in the economy and a decrease in demand for energy. 

Henry Hub Natural Gas futures are currently the third most traded physical commodity in the world, and they are included in features for similar reasons as the crude oil futures - the price movements are a good approximation of the world's economy energy demand, which in turn has great effect on the traded assets.  

Corn futures are the most traded food-related commodity in the world, and their price movements provide insight into the state of the economy in several ways. A rise in corn futures prices can indicate strong demand for corn as a feed grain for livestock, as a key ingredient in various processed food products, and as a biofuel feedstock. This strong demand could be driven by factors such as population growth, rising incomes, and government policies that support biofuels. Conversely, a fall in corn futures prices can indicate weaker demand due to factors such as a slowdown in the economy, a decrease in biofuel production, or a good crop yield. Additionally, changes can also reflect shifts in supply due to weather conditions, pest and disease outbreaks, trade policies, and other factors.

Gold is the only precious metal included in the feature space. Historically, a rise in gold prices indicated uncertainty or instability in the markets, as investors often turned to gold as a safe haven during times of turmoil. Declines in gold prices often indicated increased confidence in the stability of riskier assets such as stocks. Gold price also reflects mining production, central banks buying or selling it as a store of value, and other factors, making it a good candidate for informative features.

Copper and Aluminum are the two most demanded industrial metals, and even though their use is different, they both provide insight into activity levels in construction, transportation, and manufacturing. Growth in prices for these metals generally correlated positively with population growth, rising incomes, and government infrastructure spending. The decline in these prices was usually caused by the slowdown of industrial activity and government infrastructure spending.

Figure 4 illustrates the varied composition of our feature space, each component representing a distinct part of the economy. Feature space selection aims to reflect what traders might look at when making their trading decisions. Our goal is to cover a wide range of economic indicators, providing a clearer and more complete picture of the factors affecting the market.

Table 2 displays our list of features along with the anticipated effects of their increase and decrease. These effects are inferred from standard economic literature, as well as the intuitive understanding of market behaviors. This table aims to provide an overview of how each feature might influence market conditions, drawing from established economic principles and practical market insights.

\begin{figure}
  \centering
  \caption{Visualization of main features used for the model.}
      \makebox[\columnwidth][c]{\includegraphics[width=0.6\columnwidth]{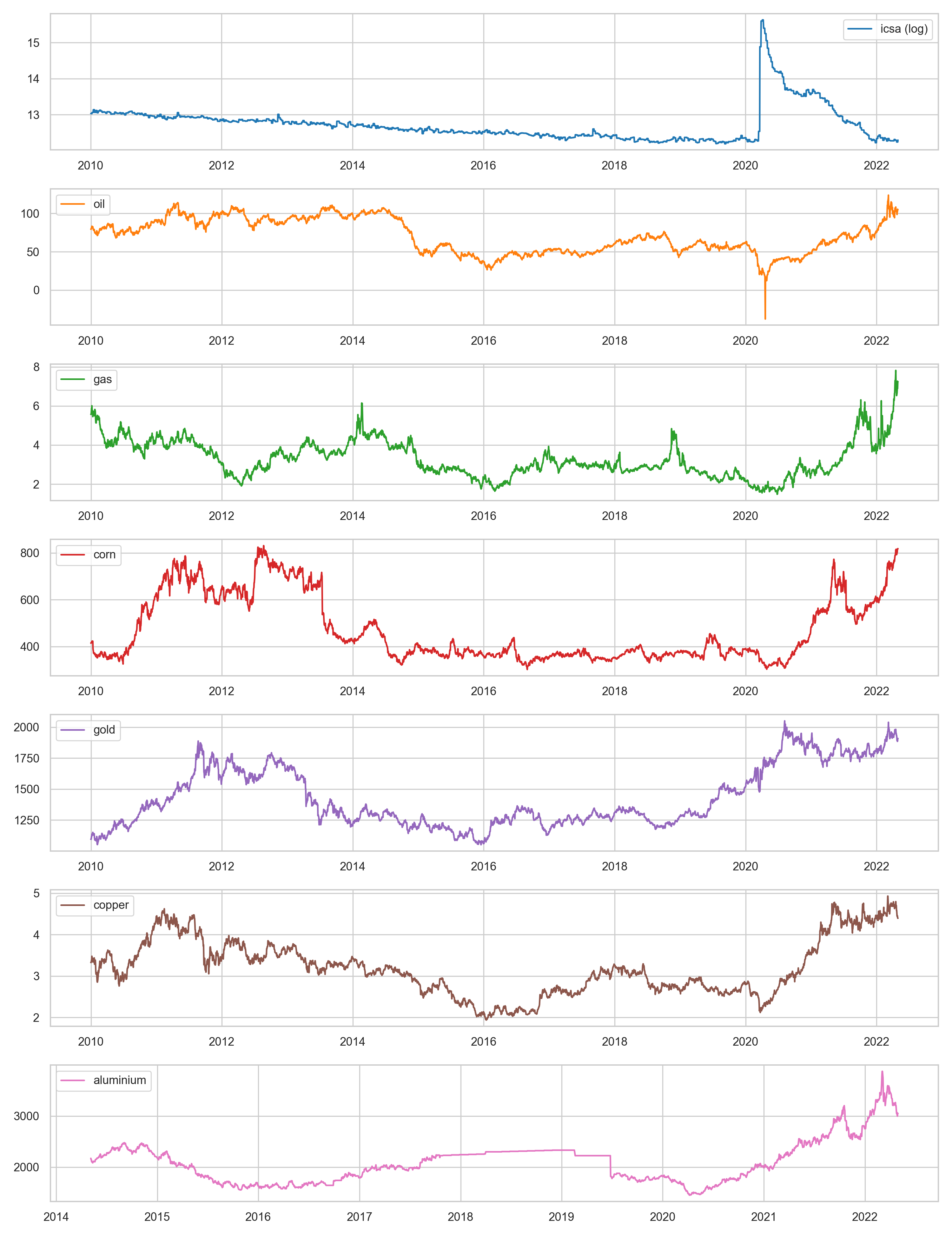}}
    \caption*{\footnotesize Source: Own Elaboration. Plotted with Matplotlib package.}
\end{figure}

\renewcommand{\arraystretch}{1.5}
\begin{table*}
\centering
\caption{Presumed impact of prices/values change on the global economy.}
{\small
\begin{tabular}{p{2cm}p{6cm}p{6cm}}
\toprule
Feature & Presumed Increase Impact & Presumed Decrease Impact \\
\midrule
ICSA & Negative: Indicates rising unemployment, potential economic slowdown & Positive: Suggests decreasing unemployment, potential economic growth \\

Oil & Mixed: Benefits oil exporters, increase costs for importers and consumers & Mixed: Lowers costs for importers and consumers, but may harm oil-exporting economies \\

Gas & Negative: Increases energy costs, affects consumer spending and production costs & Positive: Decreases energy costs, boosts consumer spending and lowers production costs \\

Corn & Negative: Raises food and feed prices, impacts food industry and inflation & Positive: Lowers food and feed prices, beneficial for food industry and inflation control \\

Gold & Mixed: Often seen as a safe haven, the increase may indicate economic uncertainty & Mixed: A decrease may reflect investor confidence, but could impact gold-producing economies \\

Copper & Positive: Suggests industrial growth and demand, often a positive economic indicator & Negative: May indicate reduced industrial activity and economic slowdown \\

Aluminium & Positive: Indicates industrial demand and growth, especially in construction and manufacturing & Negative: Could suggest a slowdown in key industries like construction and manufacturing \\
\bottomrule
\end{tabular}}
\caption*{\footnotesize Source: Own Elaboration.}
\end{table*}
\renewcommand{\arraystretch}{1}

\section{Feature Engineering}

\subsection{Motivation}
Financial time series are notorious for their low signal-to-noise ratios, a concept emphasized by Lopez de Prado (2018). This characteristic is largely due to arbitrage forces in the market. Standard stationarity transformations, such as integer differentiation, often exacerbate this issue by eliminating valuable historical memory, even though price series inherently depend on their past. Integer differentiated series, like returns, truncate this history after a limited sample window. To uncover any residual signal in these transformed series, statisticians then apply complex mathematical methods, which can lead to false discoveries. In this context, the use of ARFIMA (Autoregressive Fractionally Integrated Moving Average), models, initially proposed by Granger and Joyeux (1980), offers an insightful perspective. These models allow for fractional differentiation, providing a more nuanced approach to maintaining data stationarity while maximizing memory retention.

For inferential analysis, researchers typically transform these series into invariant processes such as returns on prices or changes in yield. This achieves stationarity but at the cost of the series' memory. While stationarity is crucial for inference, completely erasing memory is not ideal in signal processing, as memory is a key component of a model’s predictive power. Stationary models, for instance, depend on some level of memory to assess deviations in the price process from long-term expectations. The challenge, therefore, is to determine the minimal level of differentiation required to render a price series stationary while retaining as much memory as possible. Our goal is to extend the concept of returns to include stationary series where memory is not entirely discarded.

Cointegration methods have been valued for their ability to model series with memory. However, we view zero differentiation as arbitrary as 1-step differentiation. The spectrum between fully differentiated and non-differentiated series presents opportunities for fractional differentiation, a concept integral to ARFIMA models, to enhance the predictability of ML models.

Supervised learning requires stationary features because it involves mapping an unseen observation to labeled examples to predict the observation's label. Without stationarity, this mapping becomes unreliable. However, achieving stationarity alone does not ensure predictive accuracy. It is a necessary but not sufficient condition for optimal ML performance. The challenge lies in finding the right balance between achieving stationarity and retaining memory. Over-differentiation may increase stationarity but at the expense of memory, which can impair the forecasting ability of an ML algorithm. This chapter explores a methodology to address this balance, drawing on the principles of ARFIMA models and the insights of Lopez de Prado, to optimize feature engineering in financial time series analysis.

\subsection{Fractionally Differentiated Features}

DePrado(2015) introduces the concept applying ARFIMA assumptions to machine learning features - fractionally differentiated features. We consider the backshift operator $B$ applied to a time series of a feature $\{X_t\}$ such that $B^k X_t = X_{t-k}$. It follows that the difference between current and last feature's value can be expressed as $(1-B)^ X_t$. For example, $(1-B)^2 = 1-2B+B^2$, where $B^2 X_t = X_{t-2}$ so that $(1-B)^2 X_t = X_t - 2X_{t-1} + X_{t-2}$. For any positive integer $n$, it also holds that:

\begin{equation}
    (x+y)^n = \sum_{k=0}^n \binom{n}{k} x^k y^{n-k} = \sum_{k=0}^n \binom{n}{k} x^{n-k} y^{k}
\end{equation}

On the other hand for any real number $d$:

\begin{equation}
    (1+x)^d = \sum_{k=0}^\infty \binom{d}{k} x^k
\end{equation}

is the binomial series. In a model where $d$ is allowed to be a real number, the binomial series can be expanded into a series of weights which can be applied to feature values:

\begin{equation}
    \omega = \left\{ 1, -d, \frac{d(d-1)}{2!}, \frac{d(d-1)(d-1)}{3!}, ..., (-1)^k \prod_{i=0}^{k-1} \frac{d-i}{k!} \right\}
\end{equation}

The essence of fractional differencing of features is that it allows us to generalize the idea of differentiation to non-integer orders. By applying the binomial series expansion to the differencing operator, we can compute differences of any real order $d$. This means we are not limited to just taking the first, second, or nth difference, but can compute a "fractional" difference that may lie somewhere between these whole numbers. This fractional differencing can capture long-term memory in time series data while ensuring stationarity, which is crucial for many time series analysis and modeling techniques. Figure 5 displays the weights associated with each lag depending on the value on d. By adjusting the value of $d$, we can achieve a balance between removing noise and preserving meaningful information in the series.

\begin{figure}
\caption{$\omega_k$ as $k$ increases. Each line is associated with different value of $d$.} 
  \centering
      \includegraphics[width=0.6\columnwidth]{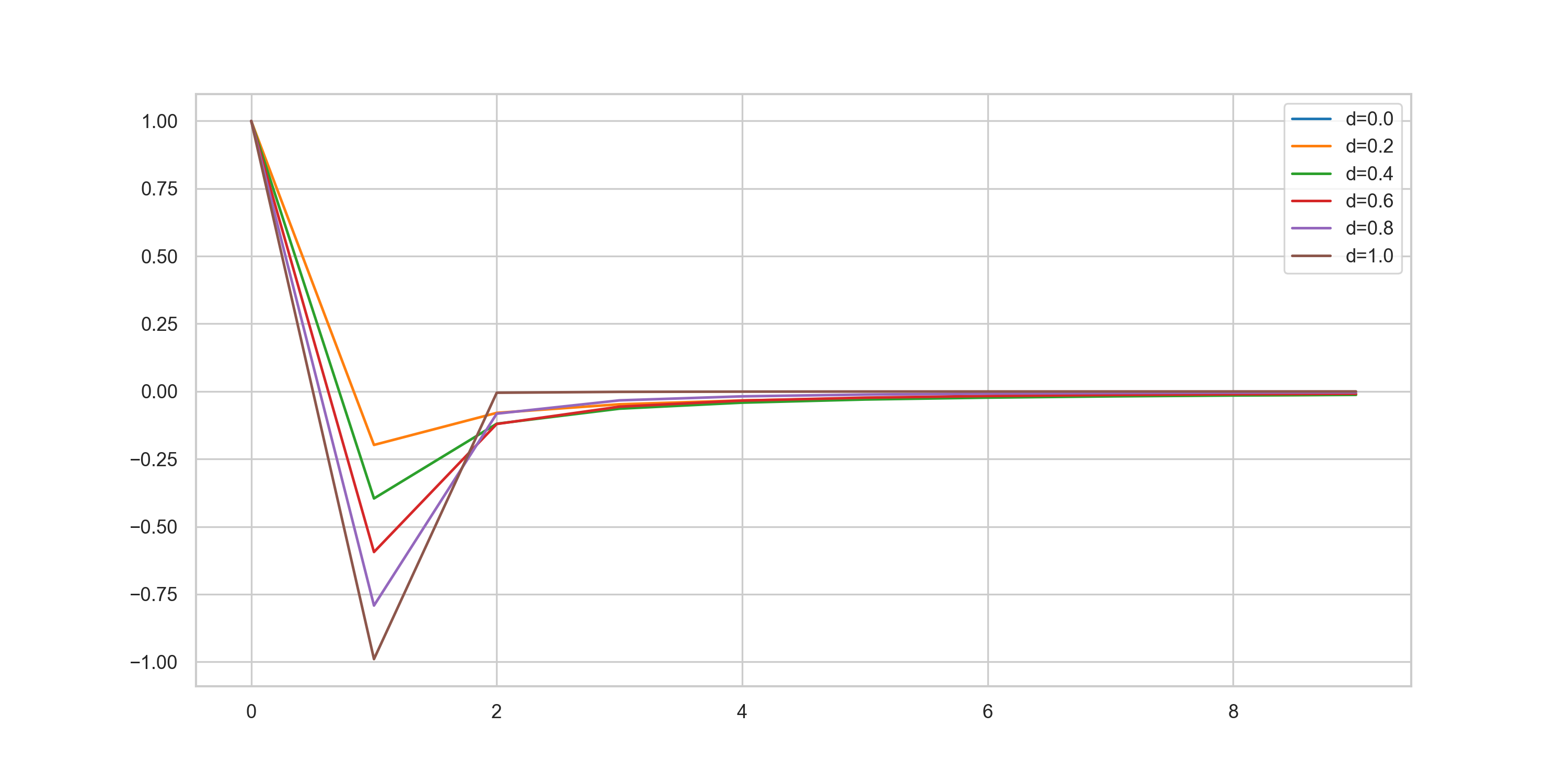}
  \caption*{\footnotesize Source: Own Elaboration based on: Marcos Lopez de Prado, Advances in Financial Machine Learning, 2018}
\end{figure}

As we are looking to implement the concept of fractionally differentiated features, a critical decision arises: determining the optimal value of $d$, the order of fractional differentiation. This value plays a pivotal role in dictating the balance between retaining memory of past values and ensuring stationarity in the time series data. While memory captures the inherent dependencies and patterns over time, stationarity facilitates robust modeling and prediction.

\subsection{Optimal Fractional Differentiation Order}

For each feature $X_t$ using the fixed-width window fractional differentiation (FFD) method allows us to determine the minimum coefficient $d$ that makes the fractionally differentiated series $X_t$ stationary (i.e. passes the Augmented Dickey-Fuller test). This coefficient $d$ signifies the memory extent that must be eliminated to attain stationarity. If $X_t$ is already stationary, $d$ = 0. For a unit root, $d < 1$, while for explosive behaviors, $d > 1$. An especially intriguing scenario is $0 < d < 1$, indicating the series is "slightly non-stationary". Here, although differentiation is essential, a complete integer differentiation might overly eliminate memory and predictive capability.

Figure 6 visualizes a concept using the ADF statistic computed on E-mini S\&P 500 futures prices, rolled forward and downsampled to daily frequency, spanning back to the contract's beginning. The x-axis represents the $d$ value associated with the ADF statistic on the right y-axis. The original series has an ADF value of $-0.3387$, while the returns series is at $-46.9114$. The ADF statistic surpasses the 95\% confidence threshold of $-2.8623$ around $d = 0.35$. The left y-axis indicates the correlation between the original and differentiated series for various $d$ values. Remarkably, the correlation remains high at $0.995$ for $d = 0.35$, suggesting successful stationarity achievement without significant memory loss. In contrast, standard integer differentiation results in a meager $0.03$ correlation with the original series, almost entirely erasing memory.
\begin{figure}
\caption{$\omega_k$ as $k$ increases. Each line is associated with a different value of $d$.} 
  \centering
      \includegraphics[width=0.6\columnwidth]{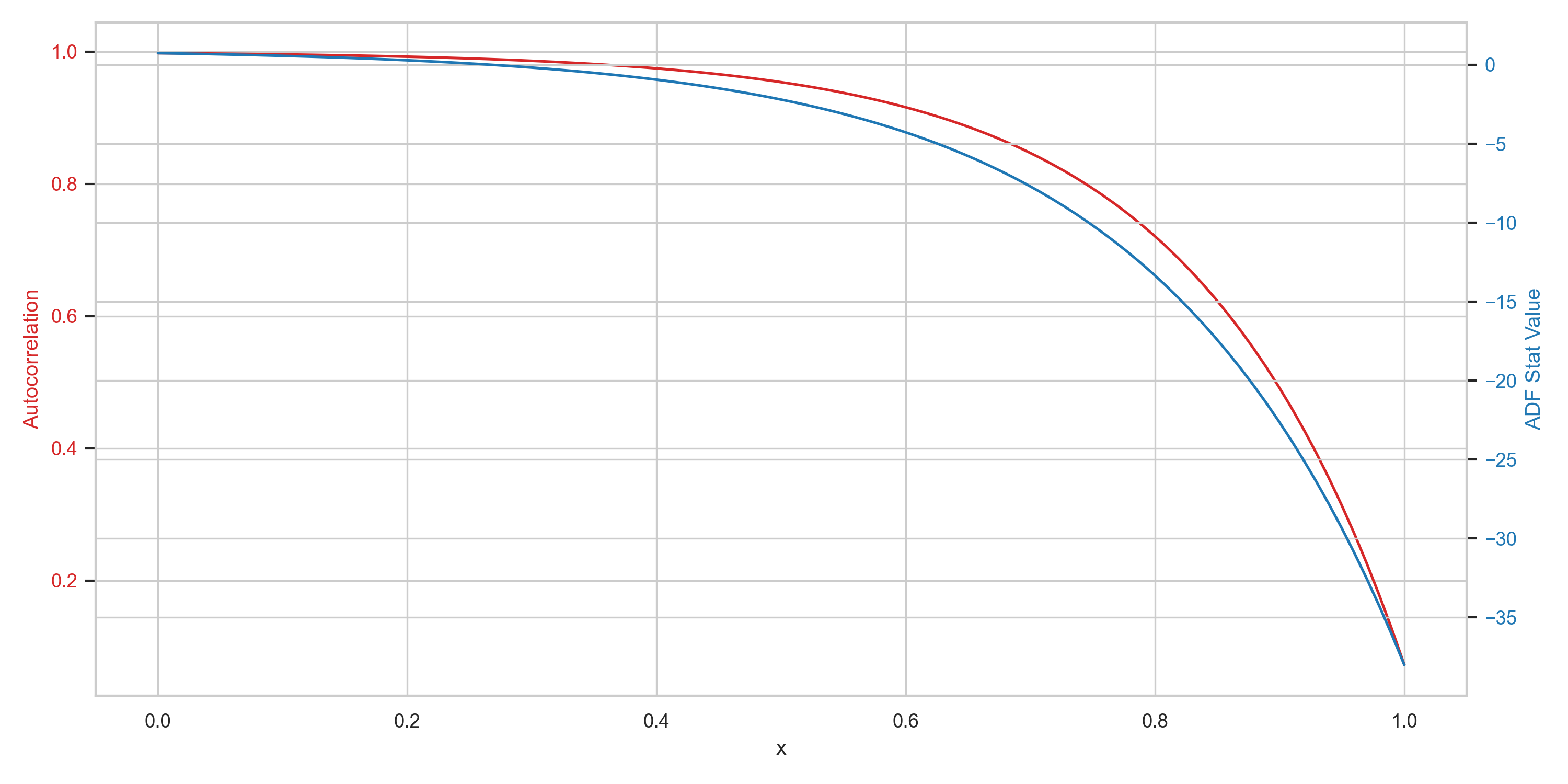}
  \caption*{\footnotesize Source: Own Elaboration based on: Marcos Lopez de Prado, Advances in Financial Machine Learning, 2018}
\end{figure}
Most finance studies lean towards integer differentiation d = 1, significantly higher than 0.35. This suggests a tendency to over-differentiate, removing more memory than required by standard econometric assumptions. Table 3 presents the results of applying fractional differentiation across a range of values, $d$, from 0.00 to 1.00 on the feature space used for the model.  Two metrics are considered for each asset at every $d$-value: the ADF p-value, which gauges the degree of stationarity, and the correlation metric, which quantifies how closely the differentiated series resembles the original series (thus serving as an indicator of memory retention). The majority of the assets achieve stationarity by a differentiation value of $d = 0.35$. Particularly, assets like CL and HH hint at stationarity even at $d = 0.10$. Conversely, more stubborn series like Bitcoin and SP500 only verge on stationarity for $d > 0.30$ (at $\alpha = 0.01$). 

Figure 7 offers a more intuitive visualization of how fractional differentiation progressively transitions the Corn price series into a series of price differences. This figure highlights the need for equilibrium between maintaining stationarity and preserving memory within the series, which typically occurs in the range between \(d=0\) and \(d=1.0\).
\begin{center}
\begin{table*}[tb]
\centering
\caption{Optimal differentiation orders for each feature used.}
{\scriptsize 

\begin{tabular}{
  l
  @{\hspace{0.35cm}}
  c@{\hspace{0.05cm}}c 
  @{\hspace{0.35cm}}
  c@{\hspace{0.05cm}}c 
  @{\hspace{0.35cm}}
  c@{\hspace{0.05cm}}c 
  @{\hspace{0.35cm}}
  c@{\hspace{0.05cm}}c 
  @{\hspace{0.35cm}}
  c@{\hspace{0.05cm}}c 
  @{\hspace{0.35cm}}
  c@{\hspace{0.05cm}}c 
  @{\hspace{0.35cm}}
  c@{\hspace{0.05cm}}c 
  @{\hspace{0.35cm}}
  c@{\hspace{0.05cm}}c 
  @{\hspace{0.35cm}}
  c@{\hspace{0.05cm}}c 
}
\toprule
& \multicolumn{2}{c}{SPX}
& \multicolumn{2}{c}{BTC}
& \multicolumn{2}{c}{EURUSD}
& \multicolumn{2}{c}{CL}
& \multicolumn{2}{c}{HH}
& \multicolumn{2}{c}{ZC}
& \multicolumn{2}{c}{GC}
& \multicolumn{2}{c}{HG}
& \multicolumn{2}{c}{ALI}
\\
d & ADF p & corr & ADF p & corr & ADF p & corr & ADF p & corr & ADF p & corr & ADF p & corr & ADF p & corr & ADF p & corr & ADF p & corr \\
\midrule
0.00 & 0.97 & 0.97 & 0.81 & 0.79 & 0.39 & 1.00 & 0.26 & 0.97 & 0.18 & 0.94 & 0.50 & 0.99 & 0.09 & 0.99 & 0.37 & 0.99 & 0.08 & 0.99 \\
0.05 & 0.83 & 0.97 & 0.89 & 0.80 & 0.19 & 1.00 & 0.06 & 0.96 & 0.07 & 0.93 & 0.28 & 0.98 & 0.08 & 0.99 & 0.17 & 0.99 & 0.03 & 1.00 \\
0.10 & 0.53 & 0.97 & 0.77 & 0.79 & 0.08 & 0.99 & \textbf{0.00} & \textbf{0.94} & \textbf{0.01} & \textbf{0.92} & 0.17 & 0.98 & 0.02 & 0.98 & 0.01 & 0.98 & 0.10 & 0.99 \\
0.15 & 0.39 & 0.97 & 0.60 & 0.79 & 0.01 & 0.97 & 0.00 & 0.91 & 0.00 & 0.89 & 0.03 & 0.96 & \textbf{0.00} & \textbf{0.96} & \textbf{0.00} & \textbf{0.96} & 0.06 & 0.98 \\
0.20 & 0.10 & 0.97 & 0.37 & 0.78 & \textbf{0.00} & \textbf{0.94} & 0.00 & 0.86 & 0.00 & 0.84 & \textbf{0.01} & \textbf{0.93} & 0.00 & 0.93 & 0.00 & 0.93 & \textbf{0.01} & \textbf{0.97} \\
0.25 & 0.01 & 0.95 & 0.13 & 0.77 & 0.00 & 0.90 & 0.00 & 0.82 & 0.00 & 0.77 & 0.00 & 0.89 & 0.00 & 0.89 & 0.00 & 0.89 & 0.00 & 0.95 \\
0.30 & \textbf{0.00} & \textbf{0.92} & 0.01 & 0.76 & 0.00 & 0.83 & 0.00 & 0.77 & 0.00 & 0.71 & 0.00 & 0.84 & 0.00 & 0.84 & 0.00 & 0.83 & 0.00 & 0.93 \\
0.35 & 0.00 & 0.89 & \textbf{0.00} & \textbf{0.74} & 0.00 & 0.77 & 0.00 & 0.71 & 0.00 & 0.64 & 0.00 & 0.79 & 0.00 & 0.79 & 0.00 & 0.77 & 0.00 & 0.90 \\
0.40 & 0.00 & 0.86 & 0.00 & 0.71 & 0.00 & 0.70 & 0.00 & 0.65 & 0.00 & 0.58 & 0.00 & 0.74 & 0.00 & 0.73 & 0.00 & 0.70 & 0.00 & 0.87 \\
0.45 & 0.00 & 0.82 & 0.00 & 0.67 & 0.00 & 0.64 & 0.00 & 0.58 & 0.00 & 0.51 & 0.00 & 0.68 & 0.00 & 0.66 & 0.00 & 0.64 & 0.00 & 0.83 \\
0.50 & 0.00 & 0.77 & 0.00 & 0.61 & 0.00 & 0.58 & 0.00 & 0.51 & 0.00 & 0.44 & 0.00 & 0.61 & 0.00 & 0.59 & 0.00 & 0.57 & 0.00 & 0.78 \\
0.55 & 0.00 & 0.70 & 0.00 & 0.56 & 0.00 & 0.51 & 0.00 & 0.43 & 0.00 & 0.38 & 0.00 & 0.54 & 0.00 & 0.52 & 0.00 & 0.49 & 0.00 & 0.73 \\
0.60 & 0.00 & 0.62 & 0.00 & 0.51 & 0.00 & 0.43 & 0.00 & 0.36 & 0.00 & 0.31 & 0.00 & 0.46 & 0.00 & 0.44 & 0.00 & 0.42 & 0.00 & 0.66 \\
0.65 & 0.00 & 0.54 & 0.00 & 0.45 & 0.00 & 0.37 & 0.00 & 0.29 & 0.00 & 0.26 & 0.00 & 0.39 & 0.00 & 0.36 & 0.00 & 0.35 & 0.00 & 0.58 \\
0.70 & 0.00 & 0.45 & 0.00 & 0.38 & 0.00 & 0.30 & 0.00 & 0.24 & 0.00 & 0.21 & 0.00 & 0.32 & 0.00 & 0.28 & 0.00 & 0.29 & 0.00 & 0.49 \\
0.75 & 0.00 & 0.37 & 0.00 & 0.32 & 0.00 & 0.24 & 0.00 & 0.19 & 0.00 & 0.18 & 0.00 & 0.26 & 0.00 & 0.23 & 0.00 & 0.23 & 0.00 & 0.41 \\
0.80 & 0.00 & 0.30 & 0.00 & 0.25 & 0.00 & 0.20 & 0.00 & 0.15 & 0.00 & 0.14 & 0.00 & 0.21 & 0.00 & 0.18 & 0.00 & 0.18 & 0.00 & 0.33 \\
0.85 & 0.00 & 0.23 & 0.00 & 0.20 & 0.00 & 0.15 & 0.00 & 0.11 & 0.00 & 0.11 & 0.00 & 0.16 & 0.00 & 0.13 & 0.00 & 0.14 & 0.00 & 0.25 \\
0.90 & 0.00 & 0.17 & 0.00 & 0.14 & 0.00 & 0.11 & 0.00 & 0.08 & 0.00 & 0.09 & 0.00 & 0.12 & 0.00 & 0.10 & 0.00 & 0.11 & 0.00 & 0.17 \\
0.95 & 0.00 & 0.11 & 0.00 & 0.09 & 0.00 & 0.08 & 0.00 & 0.05 & 0.00 & 0.06 & 0.00 & 0.08 & 0.00 & 0.06 & 0.00 & 0.08 & 0.00 & 0.10 \\
1.00 & 0.00 & 0.03 & 0.00 & 0.00 & 0.00 & 0.03 & 0.00 & 0.01 & 0.00 & 0.03 & 0.00 & 0.03 & 0.00 & 0.01 & 0.00 & 0.03 & 0.00 & -0.02 \\
\bottomrule
\end{tabular}

} 
\caption*{\footnotesize Source: Own Elaboration based on: Marcos Lopez de Prado, Advances in Financial Machine Learning, 2018, \textbf{ADF p} - p-value for Augmented Dickey-Fuller test. \textbf{corr} - time series autocorrelation coefficient. Asset symbols: \textbf{CL} - Oil, \textbf{HH} - Natural Gas, \textbf{ZC} - Corn, \textbf{GC} - Gold, \textbf{HG} - Copper, \textbf{ALI} - Aluminium.}
\end{table*}
\end{center} 

\begin{figure}
\caption{Example of different d levels impact on corn time series.} 
  \centering
      \includegraphics[width=0.6\columnwidth]{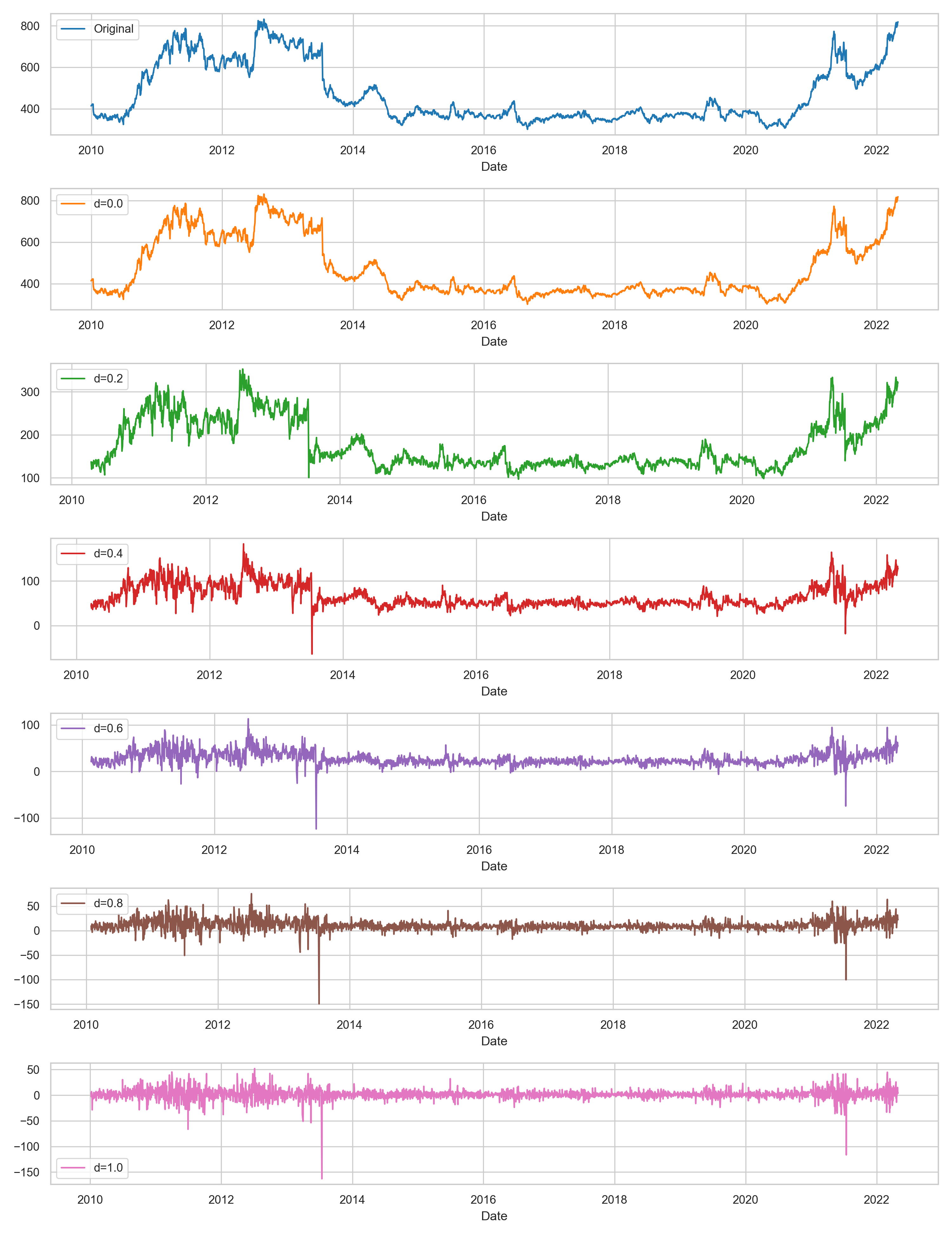}
  \caption*{\footnotesize Source: Own Elaboration based on: Marcos Lopez de Prado, Advances in Financial Machine Learning, 2018}
\end{figure}

\subsection{Implementation}

Fractional differencing emerges as a pivotal technique that attempts to balance the twin imperatives of achieving stationarity and preserving memory. Traditional integer differencing often overcorrects, eliminating more memory than necessary to meet econometric requirements, thereby diminishing the predictive power of the series. The fractional differencing method, on the other hand, enables a more nuanced transformation, allowing the retention of critical informational components within a series while ensuring its statistical propriety.

Given its advantages, fractional differencing will be applied to the features in this paper. The nature of the transformation becomes particularly useful given the unique characteristics of future data, where long memory might contain essential information for forecasting.

In terms of practical application, a walk-forward validation methodology will be adopted. This iterative process involves: 

\begin{itemize}
    \item Training Segment: Utilizing an initial segment of the data, the optimal fractional differencing parameter $d$, will be determined to achieve the balance between memory retention and stationarity.
    \item Test Segment: The identified $d$ will then be applied to the succeeding segment (or window) of the data, allowing for model evaluation.
\end{itemize}

Rolling Forward: The window then rolls forward in time, with the process iteratively repeating recalculating, and testing—ensuring that the model adapts to the evolving characteristics of the series and remains robust to varying conditions. Through this approach, we not only ensure that our models are grounded in rigorous statistical techniques, but also that they are dynamically adaptable, reflecting the evolving nature of financial markets.

In sum, the application of fractional differencing, paired with the walk-forward validation strategy, presents a robust framework for time series forecasting in futures markets, maximizing both predictive accuracy and statistical validity.

\begin{algorithm}[H]
\setstretch{1}
  \caption{Fractional Feature Differentiation in Walk-Forward Validation}
  \begin{algorithmic}[1]  
    \State Set a range of possible values for $d$ (e.g., from 0 to 1)
    \State Set significance level for ADF test (e.g., 1\%)
    \State Initiate a dictionary associating each feature with optimal $d$.
    \For{each segment pair (train, test)}
    \For{each feature}
    \State Apply fractional differencing to train segment of the feature at discreet intervals
    \State Calculate ADF test statistic and p-value for each d for a feature
    \State Choose lowest $d$ such that p-value < significance level.
    \State Save feature name and associated optimal $d$ to the dictionary
    \State Apply optimal $d$ differencing to both train and test set of the feature
    \EndFor
    \State Train the model on train segment, evaluate on test set
    \EndFor
  \end{algorithmic}
\end{algorithm}
\begin{center}
Source: Own Elaboration\\
\end{center}

\section{Triple Barrier Labelling}

\subsection{Labelling Methodology}

Most modern approaches to algorithmic trading using machine learning consist of formulating the trading as a classification problem, where the predicted class describes our position (1 - long, -1 - short, 0 - no position) in the market at a given time. In this paper, we are using the triple barrier labeling method. For specified window size $\lambda$, and maximum trade length $n$ minutes, triple barrier labelling for a given time $t$ can be expressed as: 

\begin{equation}
P_t=
    \begin{cases}
      1, & \text{if}\ \max(S_t, ..., S_{t+n}) \geq S_t \cdot (1+\lambda)\\
      -1, & \text{if}\ \min(S_t, ..., S_{t+n}) \leq S_t \cdot (1-\lambda)\\
      0, & \text{otherwise}
    \end{cases}
\end{equation}

Figure 8 represents the three cases visually. In the first case, the upper barrier was exceeded, therefore we would have preferred to be long at time $t$. In the second case, none of the horizontal barriers was exceeded, so to minimize noise in results, we ideally stay out of the market in this case. In the third case, the lower barrier was exceeded, therefore our preferred position was short. This methodology also assumes in execution, each trade has take-profit and stop-loss set at their respective $S_t \cdot (1+\lambda)$ and $S_t \cdot (1-\lambda)$ levels.\\

\begin{algorithm}[H]
\setstretch{1}
  \caption{Triple-Barrier Labeling - Simple implementation}
  \begin{algorithmic}[1]  
    \State Initialize an empty series \textit{labels} with the same index as \textit{prices}
    \For{each index \textit{idx} in \textit{prices}}
      \State Set \textit{entry\_price} to the price at \textit{idx}
      \State Calculate \textit{profit\_target} as \textit{entry\_price} \( \times (1 + \textit{profit\_taking}) \)
      \State Calculate \textit{stop\_loss\_target} as \textit{entry\_price} \( \times (1 + \textit{stop\_loss}) \)
      \State Set \textit{time\_barrier\_idx} to the minimum of \textit{idx + time\_barrier} and the last index of \textit{prices}
      \For{each price \textit{i} from \textit{idx} to \textit{time\_barrier\_idx}}
        \If{\textit{prices[i]} \(\geq\) \textit{profit\_target}}
          \State Set \textit{labels[idx]} to 1
          \State \textbf{break}
        \ElsIf{\textit{prices[i]} \(\leq\) \textit{stop\_loss\_target}}
          \State Set \textit{labels[idx]} to -1
          \State \textbf{break}
        \ElsIf{\textit{i} is equal to \textit{time\_barrier\_idx}}
          \State Set \textit{labels[idx]} to 0
          \State \textbf{break}
        \EndIf
      \EndFor
    \EndFor
    \State \Return \textit{labels}
  \end{algorithmic}
\end{algorithm}
\begin{center}
    Source: Own Elaboration
\end{center}

\begin{figure}
\caption{Triple-barrier-labelling visualization.}
  \centering
      \includegraphics[width=1\columnwidth]{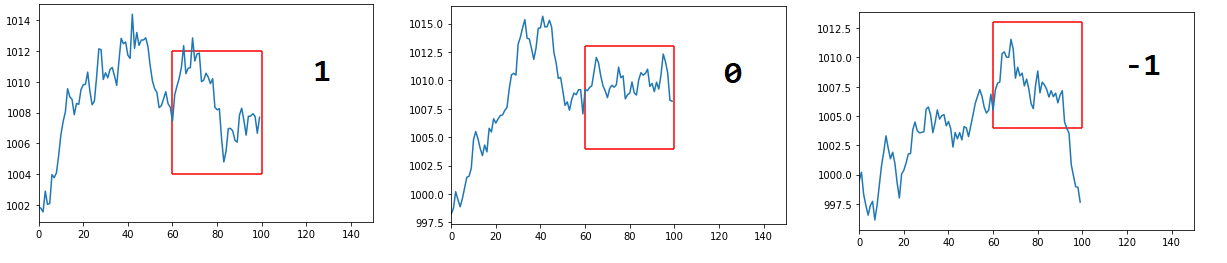}
  \caption*{\footnotesize Source: Own Elaboration based on: Marcos Lopez de Prado, Advances in Financial Machine Learning, 2018}
\end{figure}

It follows from Eq. (4) that on correct (non-zero) prediction, our return on a given trade will always be $\lambda$ (before transactional costs). We define such a case of taking the correct position as a "directly correct" prediction. The "directly incorrect" prediction ($Y_{true} = -Y_{pred}$), on the other hand, will result in a return on the trade of $-\lambda$. On "indirectly incorrect" prediction, for example, $(Y_{true} = 0, Y_{pred}=1)$ the return can be either positive or negative, depending on where the price is at $t+n$, however, will always be within $(-\lambda, \lambda)$. On predicting class 0 we do not open any position therefore our return on the trade will always be zero. Table 2 presents the return distribution for given predicted/true label combinations.

\begin{center}
Table 2. Return on a trade given classification result.\\
\vspace{0.5cm}
\begin{tabular}{|c | c c c|} 

 \hline
 Pred/True & 1 & 0 & -1 \\ [0.5ex] 
 \hline
 1 & $\lambda$ & $(-\lambda, \lambda)$ &$ -\lambda$ \\ 
 \hline
 0 & 0 & 0 & 0 \\
 \hline
 -1 & $-\lambda$ & $(-\lambda, \lambda)$ & $\lambda$ \\
 \hline
\end{tabular} \\
\vspace{0.5cm}
Source: Own Elaboration
\end{center} 

\subsection{TBL-Optimized Performance Metric}

To maximize the trading strategy performance, we must introduce an error preference mechanism. After all, missing a profitable trade (type 0 error) will have a lesser effect on our portfolio than entering an incorrect trade (type 1 error). To account for that, we cannot use the accuracy metric in our models. Instead, we have to create a novel, return-maximizing metric. 

We define \textit{directly correct count} as the number of times the model entered the correct position which resulted in the return of $\lambda$. We can similarly define \textit{directly incorrect count} as the number of times the model entered an incorrect position:

\begin{equation}
    DCC = |\{ (Y_{\text{pred}}, Y_{\text{true}}) \in S \mid Y_{\text{pred}} \neq 0 \text{ and } Y_{\text{pred}} = Y_{\text{true}} \}|
\end{equation}
\begin{equation}
    DIC = |\{ (Y_{\text{pred}}, Y_{\text{true}}) \in S \mid Y_{\text{pred}} \neq 0 \text{ and } Y_{\text{pred}} \neq Y_{\text{true}} \}|
\end{equation}

Where $|S|$ is the cardinality of set $S$. It follows from our execution assumptions that cumulative return from trades where $Y_{true} \in (-1, 1)$ can be expressed as:

\begin{equation}
    \Phi = \prod_1^{DCC} (1+\lambda) \cdot \prod_1^{DIC}(1-\lambda) = (1+\lambda)^{DCC} \cdot (1-\lambda)^{DIC} 
\end{equation}

The above equation looks like a good contender for an optimization metric. However, it is important to note that the above equation does not take into account the situation where we enter the trade and the vertical, time-based barrier is reached. We define the number of such trades as \textit{timed exit count} (TEC). We have shown that in these cases the return on the trade will be within $(-\lambda, \lambda)$, however, we cannot assume the average trade return in these situations to be 0. We can therefore introduce a preference mechanism that discourages entering such trades, which also has a much lesser "discouragement magnitude" than for directly incorrect trades. We can do that by constructing the optimization metric as if these trades on average produce a loss, however small it may be:

\begin{equation}
    \Phi = (1+\lambda)^{DCC} \cdot (1-\lambda)^{DIC} \cdot \left(1-\frac{\lambda}{\delta}\right)^{TEC}
\end{equation}

where $\delta > \lambda$. In our study, we set $\delta$ arbitrarily to 20, indicating that twenty timed exits are considered equally undesirable as one direct incorrect classification. Notably, selecting a $\delta$ value lower than $\lambda$ tends to favor trades that have historically led to timed exits. Further research is warranted to explore the potential benefits of this approach for assets characterized by consistent long-term trends, such as the S\&P 500 index.

\section{Model Training Considerations}

\subsection{Data Augmentation}

Data augmentation is a technique that has been pivotal in addressing the issues of overfitting and underrepresentation in machine learning. Originally, its use was most prominent in computer vision problems, where it significantly enhanced the performance of neural networks. Figure 9 presents exemplary data augmentation on images. In these applications, data augmentation involves making alterations to images in the training dataset to create additional training examples. These alterations can include transformations such as rotating, flipping, scaling, or altering the color balance of images. The augmented dataset thus generated presents a wider variety of scenarios for the model to learn from, which improves its ability to generalize to new, unseen images.

\begin{figure}[h]
  \centering
  \caption{Data augmentation example for computer vision problems.}
      \includegraphics[width=0.9\columnwidth]{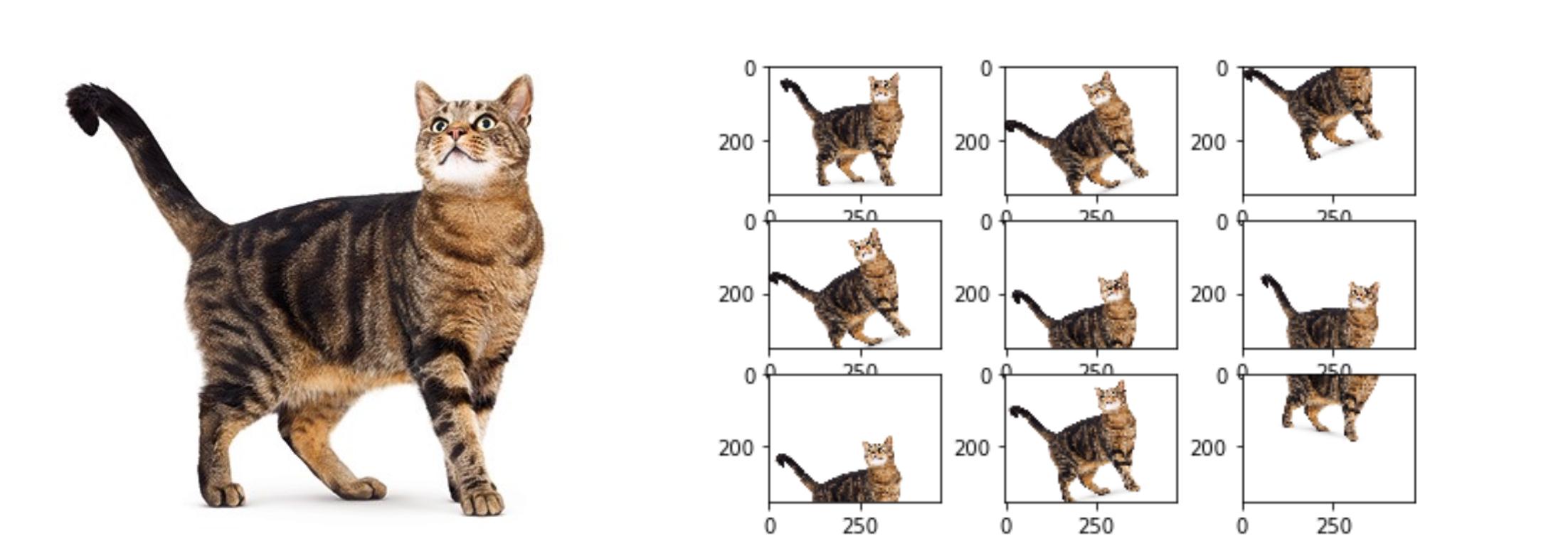}
  \caption*{\footnotesize Source: Popular techniques to prevent overfitting in neural networks, datahacker.rs, Accessed May 2022}
\end{figure}

The success of data augmentation in computer vision sparked interest in its potential applicability to other areas of machine learning. In natural language processing (NLP), for example, data augmentation might involve the paraphrasing of sentences or the use of synonyms to expand the dataset. In audio processing, it could involve varying the pitch or adding background noise to sound clips. In tabular data, techniques like feature noise injection or synthetic minority over-sampling are used to enrich the datasets.

Data augmentation has found a valuable place in the domain of time series analysis as well, which is the foundation for many algorithmic trading strategies. Time series data inherently carries the challenge of being sequential, where each point is temporally related to its predecessors and successors. In such a context, traditional data augmentation methods used in computer vision or NLP cannot be directly applied due to the risk of disrupting the time sequence, which is critical to the predictive nature of the data.

In algorithmic trading, the time series data typically consists of historical price movements, volumes, and other financial indicators that are time-dependent. To augment this type of data, the introduction of noise to features based on a fraction of historical feature volatility is an effective technique and is the basis for data augmentation in this paper. This approach preserves the temporal structure of the data while expanding the dataset. By adding noise that is a proportion of the historical volatility, one ensures that the augmented data remains realistic and within the bounds of potential market scenarios.

The noise added is typically Gaussian or drawn from a similar distribution, scaled according to the historical volatility of the feature. For example, if a particular stock has shown volatility of 2\% over a certain period, augmenting the data by adding noise with a standard deviation of 0.2\% (0.1 noise ratio) of the price feature creates new, plausible price paths for the model to learn from. This method of data augmentation helps in creating a more robust algorithmic trading strategy by forcing the model to learn not only from the historical sequence of prices but also from a range of possible price movements that could occur in real market conditions.

This technique can also be adapted to cater to multi-feature time series data where each feature may exhibit different levels of volatility. By scaling the noise for each feature individually according to its own historical volatility, the augmented data respects the relative variability of each market indicator.

The utility of adding noise based on historical feature volatility is twofold. Firstly, it helps in preventing overfitting by ensuring that the model does not learn to anticipate the exact historical sequence of events but rather the underlying patterns that govern market movements. Secondly, it increases the robustness of the algorithmic trading strategy by exposing the model to a wider variety of market conditions during the training phase, enhancing its ability to perform under different market scenarios.

In conclusion, data augmentation in time series problems, particularly for algorithmic trading, plays a critical role in model training. By judiciously adding noise to the features as a fraction of their historical volatility, one can generate a more diverse and comprehensive set of training scenarios. This approach leads to trading algorithms that are less likely to be thrown off by the inherent noise and unpredictability of financial markets and are better at generalizing from past data to future events.

\subsection{Supervised Autoencoder MLP}

An artificial neural network is a machine learning model that is inspired by the structure and function of the human brain. It is composed of layers of interconnected "neurons," which process and transmit information. Each neuron receives input from other neurons, computes the dot product of an input vector, and then sends the output to other neurons in the next layer, often with a pre-defined activation function in between.

Mathematically, a neural network can be defined as a function that maps inputs to outputs. The inputs are typically represented by a vector $x$, and the outputs are represented by a vector, $y$. Artificial neural networks are capable of approximating any continuous function, hence they are widely used in machine learning tasks.

The neural network type which is examined in detail in this paper is an autoencoder, presented in Figure 10. An autoencoder is a type of artificial neural network used to learn efficient codings of unlabeled data. The encoding is validated and refined by attempting to regenerate the input from the encoding. The autoencoder learns a representation (encoding) for a set of data, typically for dimensionality reduction, by training the network to ignore insignificant data, leading to finding the most efficient ways to compress passed data. An autoencoder consists of 3 parts: encoder, "code" (also called the bottleneck), and decoder. The encoder compresses the input and produces the code, which is the compressed, denoised data. The decoder then reconstructs the input only using the code. The metric for autoencoder performance is the similarity between reconstructed data and original input.

\begin{figure}
  \centering
  \caption{Supervised autoencoder structure.}
      \includegraphics[width=0.7\columnwidth]{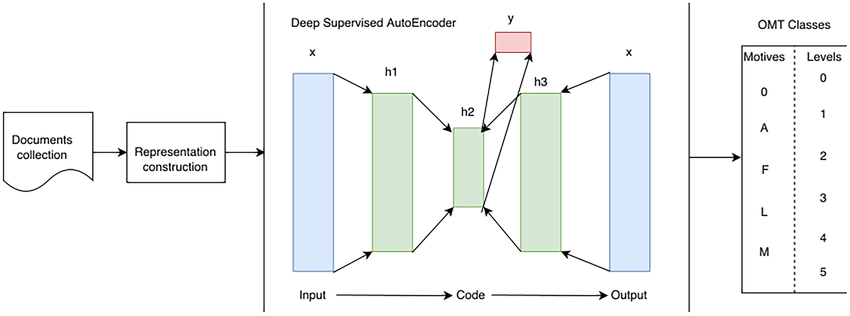}
  \caption*{\footnotesize Source: Esaú Villatoro-Tello, Shantipriya Parida, Sajit Kumar
, Petr Motlicek, Applying Attention-Based Models for Detecting Cognitive Processes and Mental Health Conditions, 2021, \textit{Cognitive Computation}}
\end{figure}

A supervised autoencoder (SAE) is a variation of an autoencoder that can be used for regression and classification tasks. In SAE, the encoded values are concatenated with the original input and used to train a supervised prediction model on provided labels. The performance metric of SAE is a combination of the accuracy of reconstruction of input data from code (unsupervised part) and the accuracy of predictions using concatenated original data and the code (supervised part). SAE models have shown to have improved generalization performance especially if the data is inherently noisy, which makes it a perfect candidate for a model in algorithmic trading problems.

\subsection{Walk-forward Validation}

In traditional validation approaches, the dataset is split into a training set and a testing set, where the model is trained on the training set and then evaluated on the testing set. However, this approach does not accurately reflect real-world scenarios, where models need to be updated and retrained regularly to adapt to the changing patterns in the data.

\begin{figure}
    \centering
    \caption{Walk-forward validation procedure. The training set, initially expanding, is limited to a 3-period length, therefore shifting instead of expanding since split 4.}
    \includegraphics[width=0.6\columnwidth]{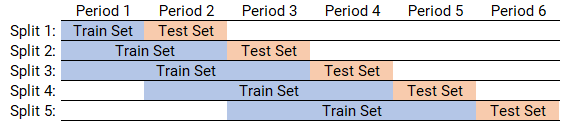}
    \caption*{\footnotesize Source: Own Elaboration. Final implementation in Python 3.10 using NumPy package}
\end{figure}

Walk-forward validation, presented in Figure 11, is a technique that involves dividing the time series dataset into multiple overlapping windows of a constant or expanding size. In each window, a model is trained on the first part of the window (train set) and evaluated on the second one (validation set). This process continues until the entire dataset has been used for training and testing the model. The window size can expand as the test size moves forward, or it can stop expanding and start shifting to ensure the model is frequently updated with the most recent data, enhancing its adaptability to new patterns. The choice of expanding versus constant size window comes down to our assumptions over the correlations in the data - do we suspect that the correlations converge to the final, "population" value, or do we expect them to shift constantly? Constant window size assumes the latter, and good performance of strategies built with this method might indicate, that the correlations change steadily enough that they can be taken advantage of.

The walk-forward validation technique has several advantages over traditional validation approaches. Firstly, it provides a more accurate estimate of the model's performance in real-world scenarios, where models need to be updated and retrained regularly. Secondly, it allows for the detection of changes in the data patterns over time, as the model is evaluated on each overlapping window. Thirdly, it ensures that the model is not overfitted to a specific portion of the dataset, as it is continuously retrained on the latest data. Finally, the constant size window fosters adaptability by requiring the model to perform well across various segments of data that reflect potential shifts in the underlying data-generating process.

The walk-forward method may also be used to tune hyperparameters. A validation period follows the in-sample and is before the out-of-sample in this scenario. The walk-forward model training with the hyperparameter adjustment procedure is analogous to the process described above, with constant window size adjustments to ensure the model remains responsive to the latest data trends.

\subsection{Construction of equity line}

In order to show how SAE-MLPs may be used in algorithmic trading, a straightforward
buy-sell trading strategy is chosen based on whether the instrument price is anticipated to
rise or fall over the next time period. For simplicity, we assume that the orders we place
will not have an effect on the market and that they are executed instantly, at the last
close price. As both S\&P 500 Index and Bitcoin markets are very liquid, this assumption
seems not far from the truth. If our model sends a “buy” signal, the strategy closes out a
short position and takes a long position. If the long position was already taken, it leaves
the position open. If the model sends a “sell” signal the algorithm takes a short position.
To calculate the cumulative unrealized P\&L the following assumptions are used:

\begin{itemize}
    \item The account is opened with \$1.000;
    \item Positions can be opened in any amount, they do not have to be full units;
    \item Transaction costs are calculated for each opening and closing of the position, which
means changing position from short to long will incur double costs. Transaction cost
for S\&P 500 Index and EUR/USD amounts to 0.005\%, for Bitcoin it is 0.1\%
\end{itemize}

\subsection{Performance Metrics}

For each strategy and asset, several indicators are computed to evaluate
profitability and performance. When evaluating portfolio performance, it is critical to consider not just the return but also the risk of the strategy. In the study, we utilize performance metrics from Michańków et al. (2022) and Ryś and Ślepaczuk (2018).

\vspace{0.5cm}
\textbf{Annualized Return Compounded}

The Annualized Return Compounded (ARC), is the constant rate of annual return
over the whole period of investment, so that:

\begin{equation}
    V(t_n) = V(t_0) \cdot (1 + ARC)^n
\end{equation}
where: \\
$V(t_0)$ - the initial value of the investment \\ 
$V(t_n)$ - the value of the investment at the end of the period \\ 
$t_n - t_0$ - number of years \\ 

\vspace{0.5cm}
\textbf{Annualized Standard Deviation (ASD)}

Volatility is a statistical indicator of the variation of returns. Most of the time, 
security is riskier the more volatile it is. Volatility may be expressed as either the standard
deviation or variation between returns from the same securities or market index. Volatility
might be easily switched to annualized values by multiplying the standard deviation of
the returns by the square root of the number of observations in a year (e.g. 252 for daily
data of the S\&P 500 Index and 365 for daily data of Bitcoin prices). In our research, we
use Annualized Standard Deviation (ASD) as a measure of volatility:

\begin{equation}
    ASD = \sqrt{\frac{1}{N-1}\sum_{t=1}^N (R_t - \hat{R})^2} \cdot \sqrt{n_{year}}
\end{equation}
where: \\
$\hat{R}$ - the average simple return (e.g. daily for daily data) of the given instrument \\ 
$R_t$ - the simple return during period t \\ 
$n_{year}$ -  number of observations in a year \\ 

\vspace{0.5cm}
\textbf{Information Ratio}

The Sharpe ratio, created by Nobel Prize winner William F. Sharpe, aids investors in
determining the return on investment relative to its risk. The ratio is the average return
over the risk-free rate for each unit of volatility or overall risk. Because we assume a
zero-rate risk-free rate, instead of Sharpe Ration we will define IR:

\begin{equation}
    IR = \frac{ARC}{ASD}
\end{equation}

\vspace{0.5cm}
\textbf{Max Drawdown (MDD)}

A portfolio’s maximum drawdown (MDD) is the largest loss that could be recorded
between a portfolio’s peak and bottom before a new high is reached. Maximum drawdown serves as a gauge for the potential loss over a certain time frame. Maximum drawdown
(MDD), a major concern for most investors is a tool used to compare the relative riskiness
of different investment strategies.

\begin{equation}
    MDD(T) = \max_{t \in [0, T]}(\max_{t \in [0, T]}V_t - V_\tau)
\end{equation}

\vspace{0.5cm}
\textbf{Max Loss Duration (MLD)}

Maximum Loss Duration (MLD) is the worst (the greatest/longest) period
between peaks that the investment has experienced. It is expressed in the number of years:

\begin{equation}
    MLD = \max \frac{m_j - m_i}{S}
\end{equation}

for which $Val(m_j) > Val(m_i)$ and $j > i. Val(m_j)$ and $Val(m_i)$ are the values of the local maximums in days $m_j$ and mi respectively. $m_j$ and mi are the numbers of days indicating
local maximums of the equity line. The scale parameter S denotes the number of trading
sessions in a year.

\vspace{0.5cm}
\textbf{Information Ratio**}

Kość et al. (2019) in their study use an additional measure to assess the effectiveness of
the strategy, which is a modification of the Information Ratio measure. This measure also
takes into account the sign of the portfolio’s rate of return and the maximum drawdown:

\begin{equation}
    IR^{**} = \frac{ARC^2 \cdot sign(ARC)}{ASD \cdot MDD}
\end{equation}

\section{Results}

\subsection{Model Description}

The difficulty of optimizing the model’s hyperparameters due to its high computational
complexity is one of the foremost challenges with neural networks. Our first approach is
to choose a set of hyperparameters using heuristic techniques and existing research, which
allows us to refit the model more than once and execute training on a rolling window. The
exact values of hyperparameters used are based on the research of Kijewski and Ślepaczuk
(2020) and presented in Table 4.

\begin{table*}[tb]
\centering
\caption{Comparison of Different Approaches}
{\scriptsize 
\begin{tabular}{
  l
  c
  c
  c
  c
}
\toprule
& Approach 1 & Approach 2 & Approach 3 & Approach 4 \\
\midrule
Activation & tanh & tanh & swish & swish \\
Loss & mse & log-loss & log-loss & log-loss \\
Epochs & 50 & 50 & 50 & 50 \\
Learning Rate & 0.01 & 0.01 & 0.01 & 0.01 \\
Hidden Layers & 1 & 1 & 1 & 1 \\
Gaussian Noise Rate & 0.00 & 0.00 & 0.05 & 0.05 \\
Problem Type & regression & classification & classification & classification \\
Model Type & Base Model & + Classification & +SAE/Denoising & +SAE/Denoising \\
& & & & +TBL \\
\bottomrule
\end{tabular}
} 
\label{tab:approaches_comparison}
\vspace{5mm}
\caption*{Source: Own Elaboration}
\end{table*}

In our study, we have implemented four different approaches to evaluate the performance of SAE-MLP models in predicting stock prices. Approach \#1 uses the walk-forward method to optimize the hyperparameters in a simple return estimating model. Given that this is a computationally intensive problem, we have utilized Bayesian Search instead of GridSearch on each window, performing 15 trials to identify the best set of hyperparameters. We have employed Mean Squared Error (MSE) as the criterion metric for selecting the best set of hyperparameters for regression issues. It is important to note that this approach requires a validation period to fine-tune the hyperparameters, which reduces the total out-of-sample duration.

For Approach \#2, we use the same neural network model as in Approach \#1, but with a classification problem where we predict the direction of the stock prices instead of forecasting returns. This requires modifying the loss function to a suitable binary classification function, such as the log-loss function.

For Approach \#3, we are retraining the model using noise augmentation and denoising with the SAE-MLP model, but keeping the simple directional classification labeling.

For Approach \#4, we use SAE-MLP noise augmentation and denoising, but this time with a classification problem where we use the aforementioned triple-barrier labeling. We also use the performance metric $\Psi$ mentioned in the labeling section of this paper.

It is worth mentioning that we have applied all 4 approaches for all three assets. Moreover, we have derived results for equally-weighted portfolio for each approach. Thus, for each approach, we have produced 16 equity lines in total (4 approaches x 3 assets) + 4 "portfolio" lines.

\subsection{Approach 1 - Regression Next-Close Forecasting}

\begin{figure}[t]
\centering 

\begin{minipage}{0.45\textwidth}
    \centering
    \caption{Cumulative returns for Approach 1 with 5-minute bar frequency compared to buy and hold.}
    \includegraphics[width=\textwidth]{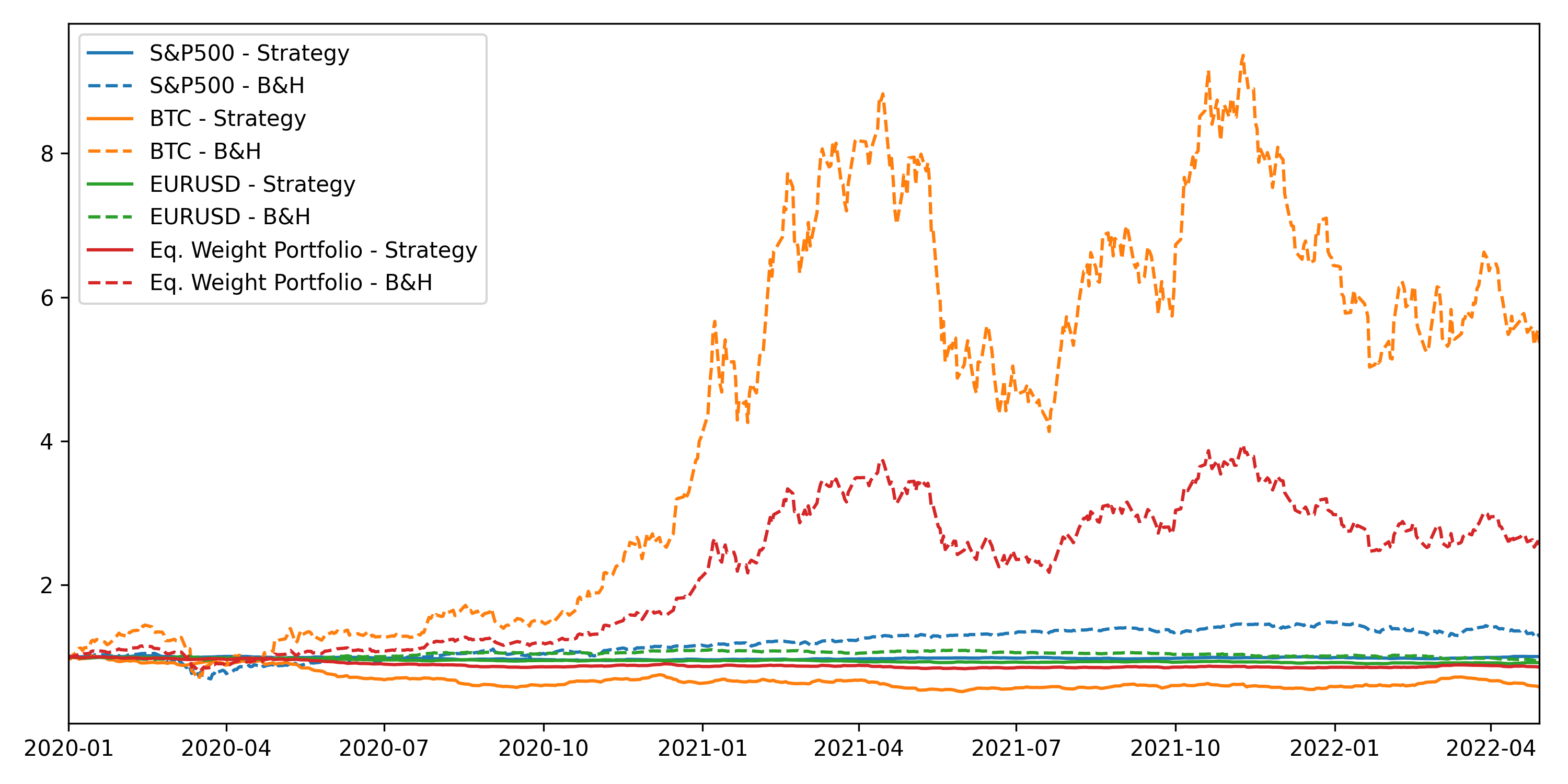} 
    \caption*{\footnotesize Source: Own Elaboration. Out-of-sample performance between 01.01.2020 and 30.04.2022, own backtesting implementation.}
\end{minipage}\hfill 
\begin{minipage}{0.45\textwidth}
    \centering
    \setstretch{1.1}
    \captionof{table}{Performance metrics: Approach 1 with 5-minute bar frequency (top) compared to buy and hold (bottom).} 
    {\scriptsize 
    \begin{tabular}{lcccc}
    \toprule
    Approach 1 - 5min & SP500 & BTCUSD & EURUSD & Eq. Weighted \\
    \midrule
    Cumulative Return & 0.57\% & -41.44\% & -11.08\% & -17.32\% \\
    Annual Return & 0.23\% & -19.92\% & -4.76\% & -7.59\% \\
    Annualized Std & 4.65\% & 24.18\% & 4.56\% & 6.98\% \\
    Information Ratio & 0.05 & -0.82 & -1.04 & -1.09 \\
    Max Drawdown & 9.51\% & 54.44\% & 13.30\% & 21.72\% \\
    MLD & 568 days & 596 days & 606 days & 596 days \\
    IR** & 0.0 & -0.3 & -0.37 & -0.38 \\
    \bottomrule
    \end{tabular}
    \begin{tabular}{lcccc}
    \toprule
    Buy And Hold & SP500 & BTCUSD & EURUSD & Eq. Weighted \\
    \midrule
    Cumulative Return & 26.83\% & 452.57\% & -5.58\% & 157.94\% \\
    Annual Return & 10.76\% & 108.57\% & -2.44\% & 50.30\% \\
    Annualized Std & 25.58\% & 69.97\% & 6.93\% & 45.38\% \\
    Information Ratio & 0.42 & 1.55 & -0.35 & 1.11 \\
    Max Drawdown & 38.25\% & 67.04\% & 15.78\% & 49.95 \% \\
    MLD & 125 days & 129 days & 331 days & 129 days \\
    IR** & 0.12 & 2.51 & -0.05 & 1.12 \\
    \bottomrule
    \end{tabular}
    }
    \caption*{\footnotesize Source: Own Elaboration. Out-of-sample performance between 01.01.2020 and 30.04.2022, own backtesting implementation.} 
\end{minipage}
\end{figure}

In analyzing the performance of Approach 1 utilizing a 5-minute bar frequency (Figure 13, Table 5), the results depicted mostly negative performance across assets. For the S\&P 500, a modest cumulative return of 0.57\% was achieved, with an information ratio of 0.05, suggesting weak performance per unit of risk. This was contrasted by a significant downturn in Bitcoin, which experienced a substantial cumulative loss of 41.44\%, culminating in an information ratio of -0.82. The EUR/USD currency pair also underperformed with an 11.08\% decrease in value and an information ratio of -1.04.

The $IR^{**}$, which adjusts for drawdown depth, was zero for the SP500, suggesting a neutral performance after accounting for the maximum drawdown of 9.51\%. In contrast, Bitcoin and EUR/USD, as well as the portfolio, showed negative $IR^{**}$s of -0.3, -0.37, and -0.38, respectively, underscoring the compounded effect of both underperformance and significant drawdowns, with the maximum drawdowns reaching over 50\% for Bitcoin, 13.30\% for EUR/USD, and 21.72\% for the portfolio.

\begin{figure}[t]
\begin{minipage}{0.45\textwidth}
    \centering
    \caption{Cumulative returns for Approach 1 with 15-minute bar frequency compared to buy and hold.}
    \includegraphics[width=\linewidth]{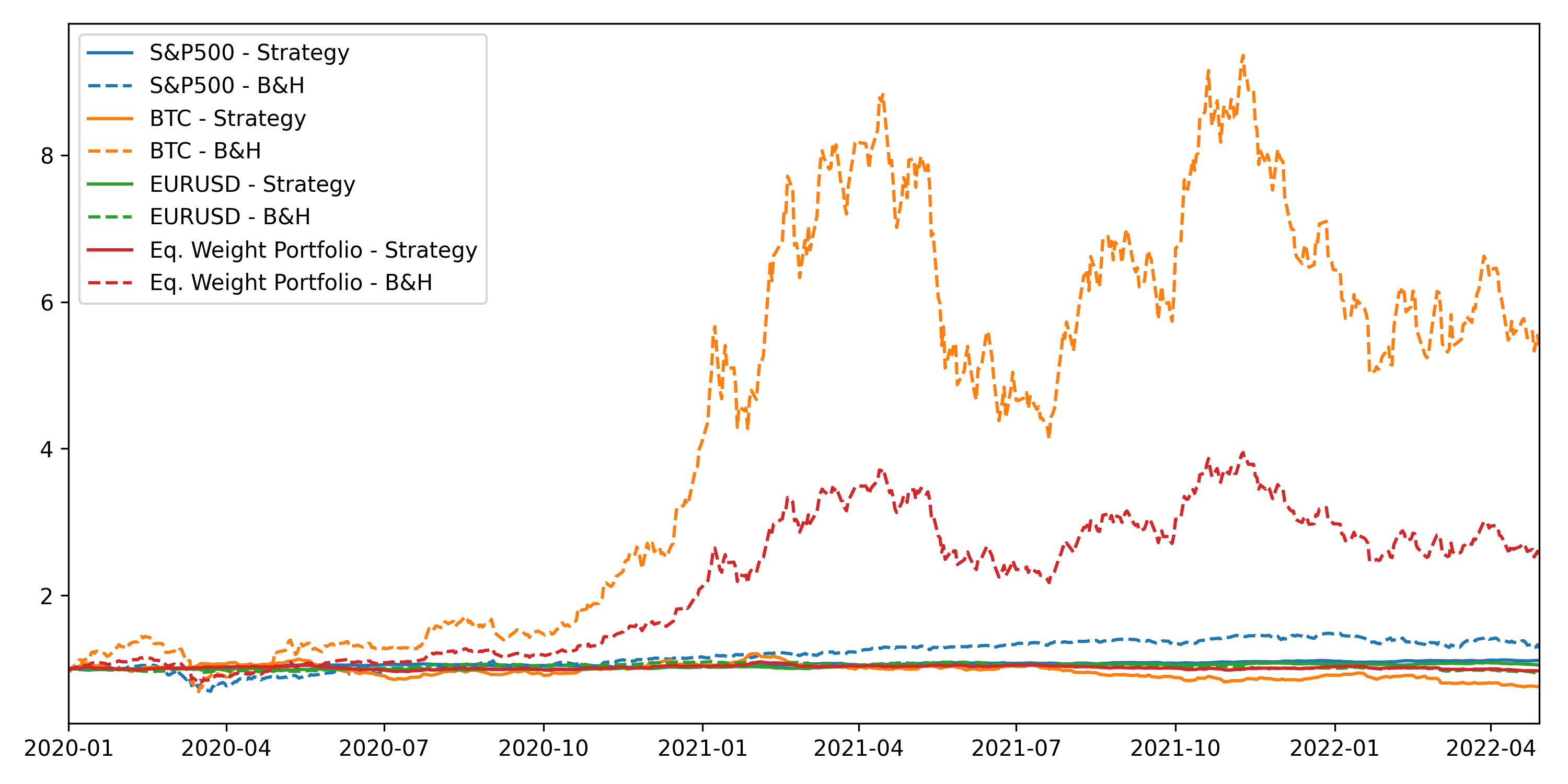}
    \label{fig:approach1_15min}
    \vspace{0.5cm} 
    \caption*{\footnotesize Source: Own Elaboration. Out-of-sample performance between 01.01.2020 and 30.04.2022, own backtesting implementation.}
\end{minipage}\hfill 
\begin{minipage}{0.45\textwidth}
    \centering
    \setstretch{1.1}
    \captionof{table}{Performance metrics: Approach 1 with 15-minute bar frequency (top) compared to buy and hold (bottom).}
    {\scriptsize 
    \begin{tabular}{lcccc}
    \toprule
    Approach 1 - 15min & SP500 & BTCUSD & EURUSD & Eq. Weighted \\
    \midrule
    Cumulative Return & 11.00\% & -25.29\% & 5.52\% & -2.92\% \\
    Annual Return & 4.43\% & -11.40\% & 2.26\% & -1.22\% \\
    Annualized Std & 4.61\% & 22.62\% & 4.98\% & 7.40\% \\
    Information Ratio & 0.96 & -0.50 & 0.45 & -0.17 \\
    Max Drawdown & 5.42\% & 47.17\% & 3.74\% & 12.72\% \\
    MLD & 154 days & 314 days & 155 days & 314 days \\
    IR** & 0.79 & -0.12 & 0.27 & -0.02 \\
    \midrule
    Buy And Hold & SP500 & BTCUSD & EURUSD & Eq. Weighted \\
    \midrule
    Cumulative Return & 26.83\% & 452.57\% & -5.58\% & 157.94\% \\
    Annual Return & 10.76\% & 108.57\% & -2.44\% & 50.30\% \\
    Annualized Std & 25.58\% & 69.97\% & 6.93\% & 45.38\% \\
    Information Ratio & 0.42 & 1.55 & -0.35 & 1.11 \\
    Max Drawdown & 38.25\% & 67.04\% & 15.78\% & 49.95 \% \\
    MLD & 125 days & 129 days & 331 days & 129 days \\
    IR** & 0.12 & 2.51 & -0.05 &1.12 \\
    \bottomrule
    \end{tabular}
    }
    \caption*{\footnotesize Source: Own Elaboration. Out-of-sample performance between 01.01.2020 and 30.04.2022, own backtesting implementation.} 
    \label{tab:approach1_15min}
\end{minipage}
\end{figure}

Approach 1 with a 15-minute bar frequency (Figure 14, Table 6) presented a slightly improved, although still differentiated outcome across the asset classes analyzed. The SP500 yielded a positive cumulative return of 11.00\% and an impressive information ratio of 0.96. This was complemented by a robust $IR^{**}$ of 0.79 after adjusting for drawdown, which was relatively shallow at 5.42\%. Bitcoin's performance remained in negative territory with a significant cumulative loss of 25.29\%, a negative information ratio of -0.50, and a $IR^{**}$ of -0.12. The EUR/USD pair displayed a positive turnaround with a cumulative return of 5.52\% and an information ratio of 0.45, with a notably low maximum drawdown of 3.74\%, resulting in a positive $IR^{**}$ of 0.27. The equally-weighted portfolio's results were mixed, with a small cumulative loss of -2.92\%, a slightly negative information ratio of -0.17, and a $IR^{**}$ just into the negative at -0.02, suggesting a neutral performance when accounting for drawdown severity, which was notably less at 12.72\%. The duration of maximum drawdown was significantly reduced across the assets, particularly for the SP500 and EUR/USD, indicating a more resilient strategy over this interval.

\begin{figure}[t]
\begin{minipage}{0.45\textwidth}
    \centering
    \caption{Cumulative returns for Approach 1 with 30-minute bar frequency compared to buy and hold.}
    \includegraphics[width=\linewidth]{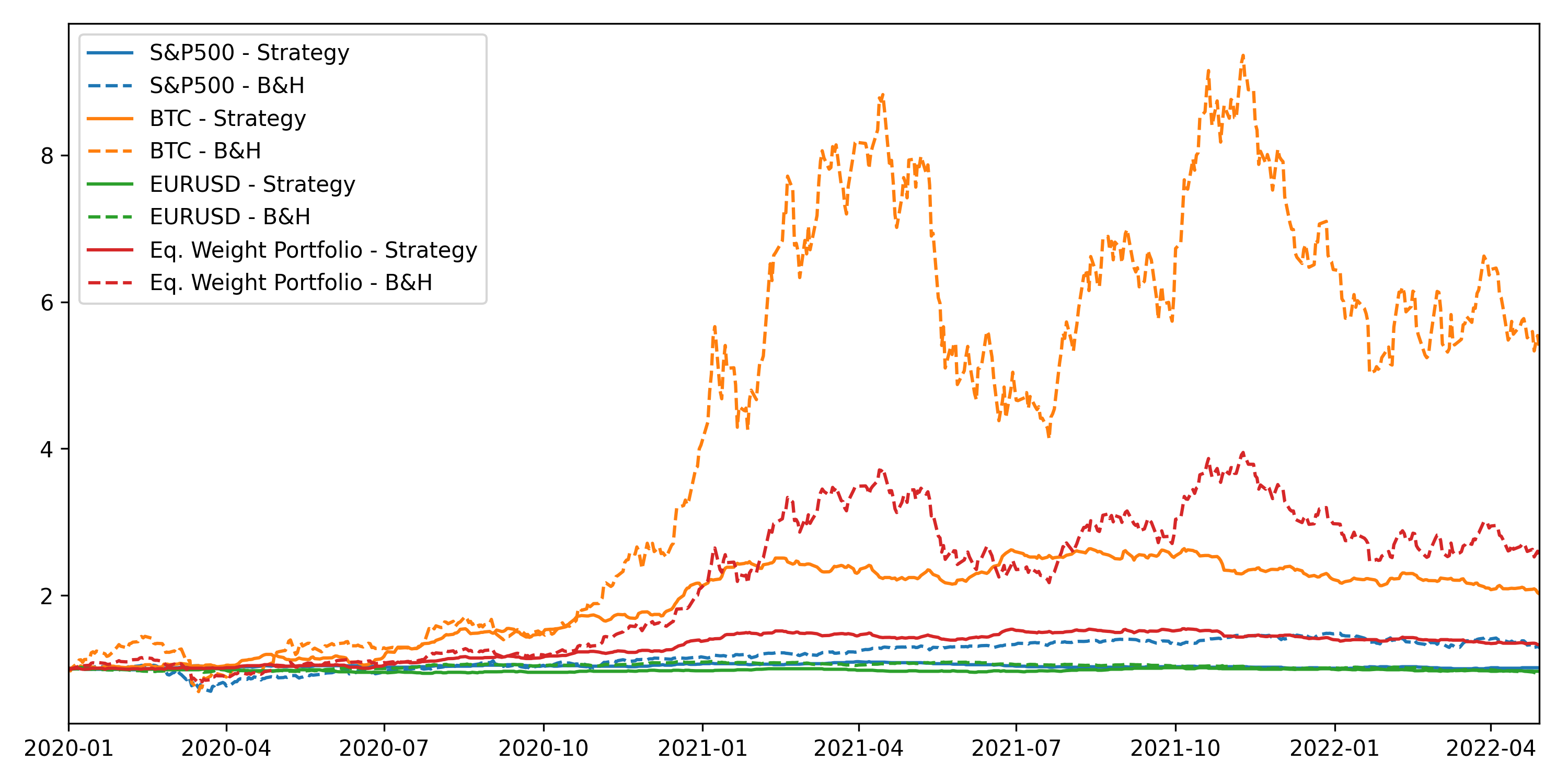}
    \label{fig:approach1_30min}
    \vspace{0.5cm} 
    \caption*{\footnotesize Source: Own Elaboration. Out-of-sample performance between 01.01.2020 and 30.04.2022, own backtesting implementation.}
\end{minipage}\hfill 
\begin{minipage}{0.45\textwidth}
    \centering
    \setstretch{1.1}
    \captionof{table}{Performance metrics: Approach 1 with 30-minute bar frequency (top) compared to buy and hold (bottom).}
    {\scriptsize 
    \begin{tabular}{lcccc}
    \toprule
    Approach 1 - 30min & SP500 & BTCUSD & EURUSD & Eq. Weighted \\
    \midrule
    Cumulative Return & 1.14\% & 98.85\% & -3.52\% & 32.16\% \\
    Annual Return & 0.47\% & 33.02\% & -1.48\% & 12.27\% \\
    Annualized Std & 4.69\% & 23.47\% & 4.53\% & 11.43\% \\
    Information Ratio & 0.10 & 1.41 & -0.33 & 1.07 \\
    Max Drawdown & 9.67\% & 62.65\% & 8.41\% & 22.78\% \\
    MLD & 265 days & 180 days & 583 days & 144 days \\
    IR** & 0.0 & 0.74 & -0.06 & 0.58 \\
    \midrule
    Buy And Hold & SP500 & BTCUSD & EURUSD & Eq. Weighted \\
    \midrule
    Cumulative Return & 26.83\% & 452.57\% & -5.58\% & 157.94\% \\
    Annual Return & 10.76\% & 108.57\% & -2.44\% & 50.30\% \\
    Annualized Std & 25.58\% & 69.97\% & 6.93\% & 45.38\% \\
    Information Ratio & 0.42 & 1.55 & -0.35 & 1.11 \\
    Max Drawdown & 38.25\% & 67.04\% & 15.78\% & 49.95 \% \\
    MLD & 125 days & 129 days & 331 days & 129 days \\
    IR** & 0.12 & 2.51 & -0.05 &1.12 \\
    \bottomrule
    \end{tabular}
    }
    \caption*{\footnotesize Source: Own Elaboration. Out-of-sample performance between 01.01.2020 and 30.04.2022, own backtesting implementation.} 
    \label{tab:approach1_30min}
\end{minipage}
\end{figure}

For approach 1, with 30-minute bars (Figure 15, Table 7), SP500 delivered a minimal cumulative return of 1.14\% with a low information ratio of 0.10. The strategy's resilience is weak as the $IR^{**}$ settled at 0.0 after accounting for the drawdown. Bitcoin, however, stood out with a remarkable cumulative return of 98.85\%, translating into an information ratio of 1.41. This performance is emphasized by a $IR^{**}$ of 0.74 after accounting for the maximum drawdown of 62.65\%. The EUR/USD pair, contrarily, faced a cumulative return of -3.52\% and a negative information ratio of -0.33. The $IR^{**}$ of -0.06 indicates a slight underperformance after factoring in the drawdown extent. The equally weighted portfolio showcased a strong cumulative return of 32.16\% and an information ratio of 1.07, suggesting effective diversification benefits. The portfolio's $IR^{**}$ of 0.58 highlights its resilience, even in light of a 22.78\% maximum drawdown. The duration of the maximum drawdown showed improvement for Bitcoin at only 180 days, while the portfolio's drawdown period was notably short at 144 days, signifying a quicker recovery from losses.

\begin{figure}[t]
\begin{minipage}{0.45\textwidth}
    \centering
    \caption{Cumulative returns for Approach 2 with 5-minute bar frequency compared to buy and hold.}
    \includegraphics[width=\linewidth]{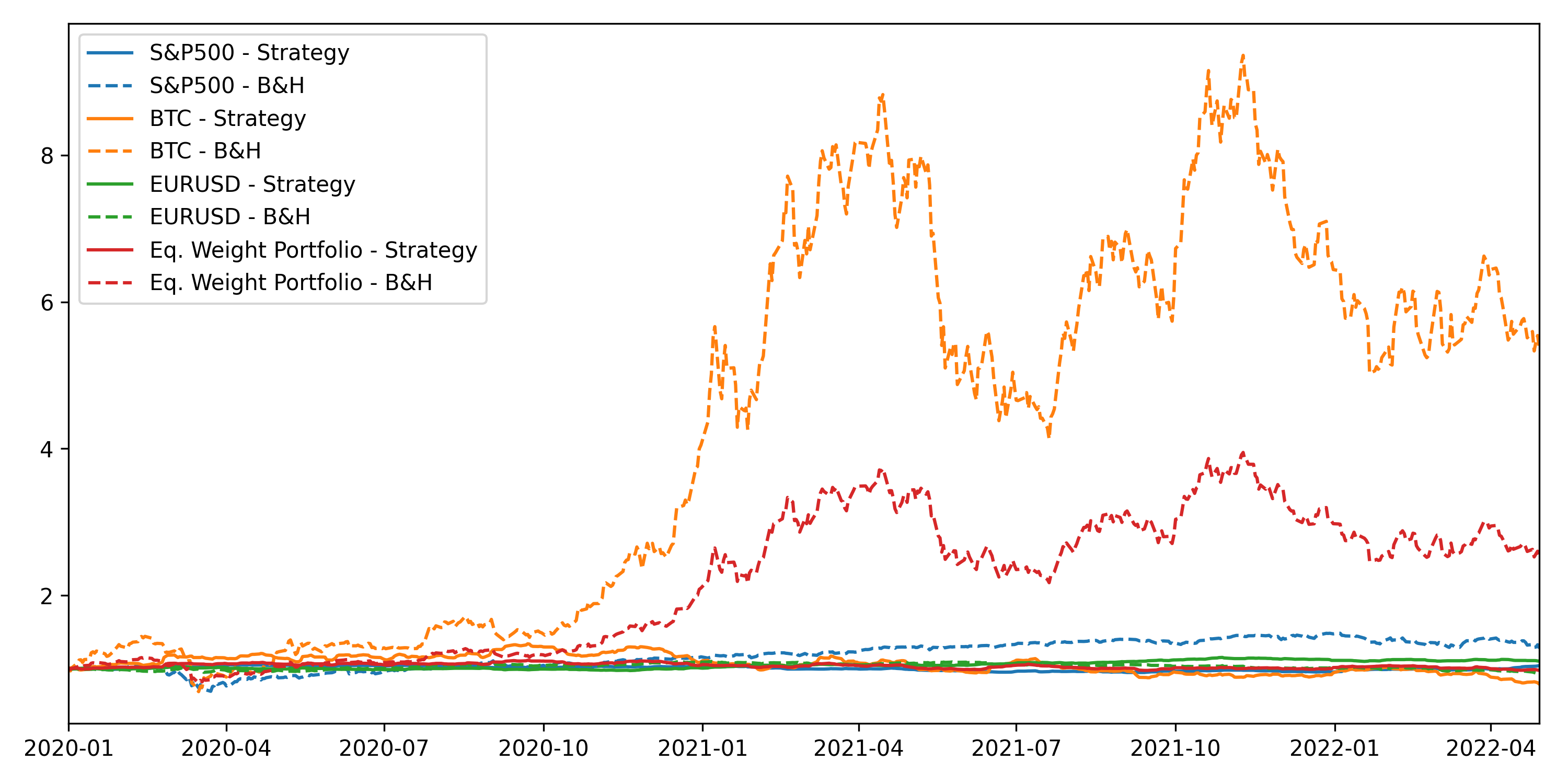}
    \label{fig:approach2_5min}
    \vspace{0.5cm} 
    \caption*{\footnotesize Source: Own Elaboration. Out-of-sample performance between 01.01.2020 and 30.04.2022, own backtesting implementation.}
\end{minipage}\hfill 
\begin{minipage}{0.45\textwidth}
    \centering
    \setstretch{1.1}
    \captionof{table}{Performance metrics: Approach 2 with 5-minute bar frequency (top) compared to buy and hold (bottom).}
    {\scriptsize 
    \begin{tabular}{lcccc}
    \toprule
    Approach 2 - 5min & SP500 & BTCUSD & EURUSD & Eq. Weighted \\
    \midrule
    Cumulative Return & 5.41\% & -21.72\% & 9.84\% & -2.16\% \\
    Annual Return & 2.21\% & -9.67\% & 3.97\% & -0.90\% \\
    Annualized Std & 4.77\% & 23.22\% & 4.82\% & 8.29\% \\
    Information Ratio & 0.46 & -0.42 & 0.82 & -0.11 \\
    Max Drawdown & 10.31\% & 56.07\% & 5.64\% & 14.06\% \\
    MLD & 532 days & 422 days & 215 days & 422 days \\
    IR** & 0.1 & -0.07 & 0.58 & -0.01 \\
    \midrule
    Buy And Hold & SP500 & BTCUSD & EURUSD & Eq. Weighted \\
    \midrule
    Cumulative Return & 26.83\% & 452.57\% & -5.58\% & 157.94\% \\
    Annual Return & 10.76\% & 108.57\% & -2.44\% & 50.30\% \\
    Annualized Std & 25.58\% & 69.97\% & 6.93\% & 45.38\% \\
    Information Ratio & 0.42 & 1.55 & -0.35 & 1.11 \\
    Max Drawdown & 38.25\% & 67.04\% & 15.78\% & 49.95 \% \\
    MLD & 125 days & 129 days & 331 days & 129 days \\
    IR** & 0.12 & 2.51 & -0.05 &1.12 \\
    \bottomrule
    \end{tabular}
    }
    \caption*{\footnotesize Source: Own Elaboration. Out-of-sample performance between 01.01.2020 and 30.04.2022, own backtesting implementation.} 
    \label{tab:approach2_5min}
\end{minipage}
\end{figure}

\subsection{Approach 2 - Directional classifier}

For approach 2, 5-minute bars (Figure 16, Table 8), SP500 yielded a cumulative return of 5.41\% on the test set, reflecting an information ratio of 0.46. However, the $IR^{**}$ falls to 0.1 when accounting for a maximum drawdown of 10.31\%. Bitcoin faced a notable downturn with a cumulative loss of 21.72\% and an information ratio of -0.42, though the $IR^{**}$ at -0.07 implies a slightly less negative outlook when considering the drawdown. The EUR/USD pair exhibited strong performance with a cumulative return of 9.84\% and an information ratio of 0.82, indicating a solid performance per unit of risk. This is reinforced by a high $IR^{**}$ of 0.58, despite a maximum drawdown of 5.64\%, pointing to efficient risk-adjusted returns. The equally-weighted portfolio, however, resulted in a marginal cumulative loss of -2.16\%, an information ratio of -0.11, and a nearly neutral $IR^{**}$ of -0.01, reflecting an overall balanced but slightly negative performance after factoring in the maximum drawdown of 14.06\%.

\begin{figure}[t]
\begin{minipage}{0.45\textwidth}
    \centering
    \caption{Cumulative returns for Approach 2 with 15-minute bar frequency compared to buy and hold.}
    \includegraphics[width=\linewidth]{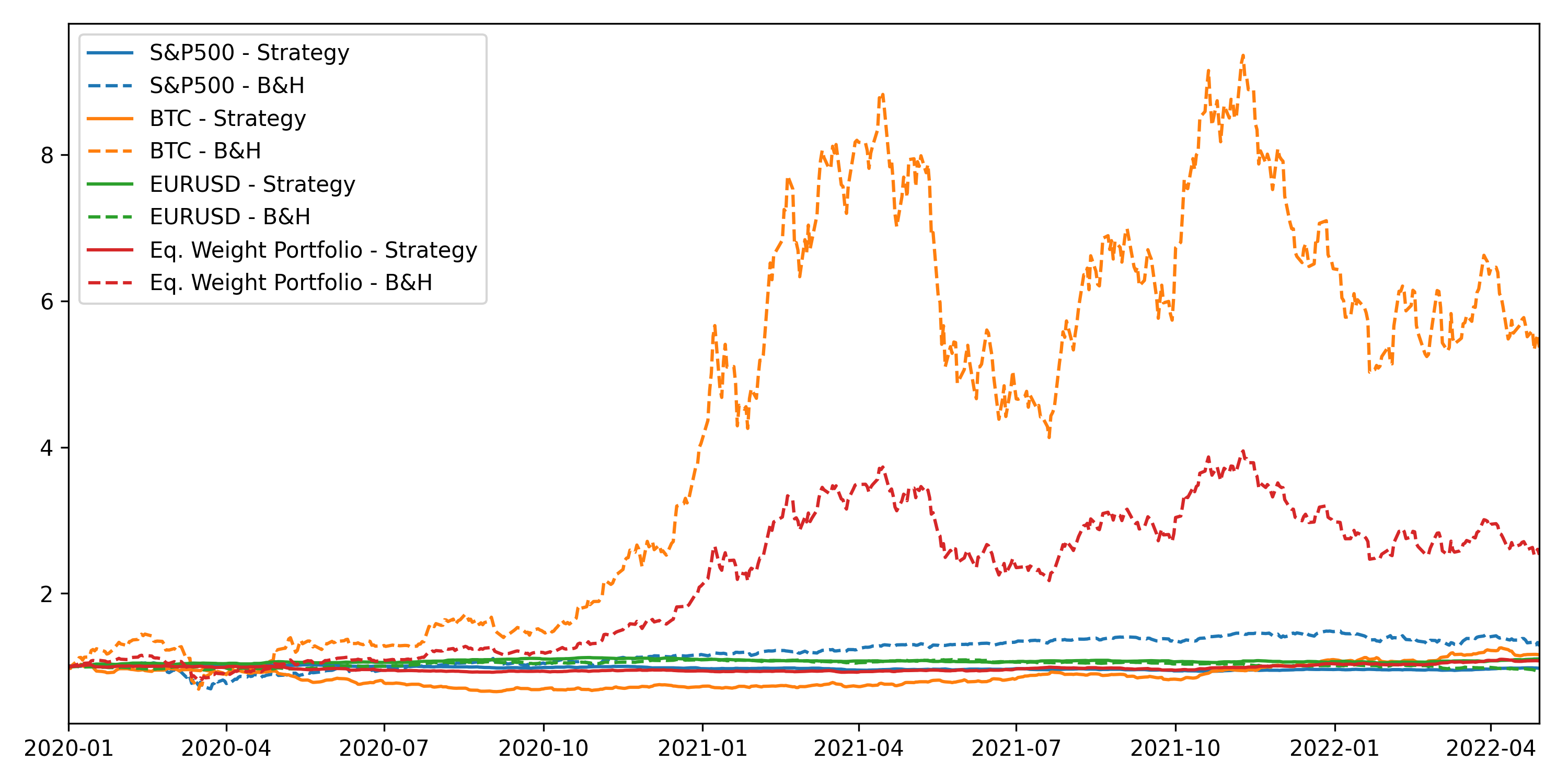}
    \label{fig:approach2_15min}
    \vspace{0.5cm} 
    \caption*{\footnotesize Source: Own Elaboration. Out-of-sample performance between 01.01.2020 and 30.04.2022, own backtesting implementation.}
\end{minipage}\hfill 
\begin{minipage}{0.45\textwidth}
    \centering
    \setstretch{1.1}
    \captionof{table}{Performance metrics: Approach 2 with 15-minute bar frequency (top) compared to buy and hold (bottom).}
    {\scriptsize 
    \begin{tabular}{lcccc}
    \toprule
    Approach 2 - 15min & SP500 & BTCUSD & EURUSD & Eq. Weighted \\
    \midrule
    Cumulative Return & -3.08\% & 19.92\% & 11.73\% & 9.52\% \\
    Annual Return & -1.29\% & 7.83\% & 4.71\% & 3.85\% \\
    Annualized Std & 4.73\% & 22.99\% & 4.79\% & 7.32\% \\
    Information Ratio & -0.27 & 0.34 & 0.98 & 0.53 \\
    Max Drawdown & 10.72\% & 41.52\% & 8.38\% & 12.60\% \\
    MLD & 479 days & 505 days & 382 days & 500 days \\
    IR** & -0.03 & 0.06 & 0.55 & 0.16 \\
    \midrule
    Buy And Hold & SP500 & BTCUSD & EURUSD & Eq. Weighted \\
    \midrule
    Cumulative Return & 26.83\% & 452.57\% & -5.58\% & 157.94\% \\
    Annual Return & 10.76\% & 108.57\% & -2.44\% & 50.30\% \\
    Annualized Std & 25.58\% & 69.97\% & 6.93\% & 45.38\% \\
    Information Ratio & 0.42 & 1.55 & -0.35 & 1.11 \\
    Max Drawdown & 38.25\% & 67.04\% & 15.78\% & 49.95 \% \\
    MLD & 125 days & 129 days & 331 days & 129 days \\
    IR** & 0.12 & 2.51 & -0.05 &1.12 \\
    \bottomrule
    \end{tabular}
    }
    \caption*{\footnotesize Source: Own Elaboration. Out-of-sample performance between 01.01.2020 and 30.04.2022, own backtesting implementation.} 
    \label{tab:approach2_15min}
\end{minipage}
\end{figure}

\begin{figure}[t]
\begin{minipage}{0.45\textwidth}
    \centering
    \caption{Cumulative returns for Approach 2 with 30-minute bar frequency compared to buy and hold.}
    \includegraphics[width=\linewidth]{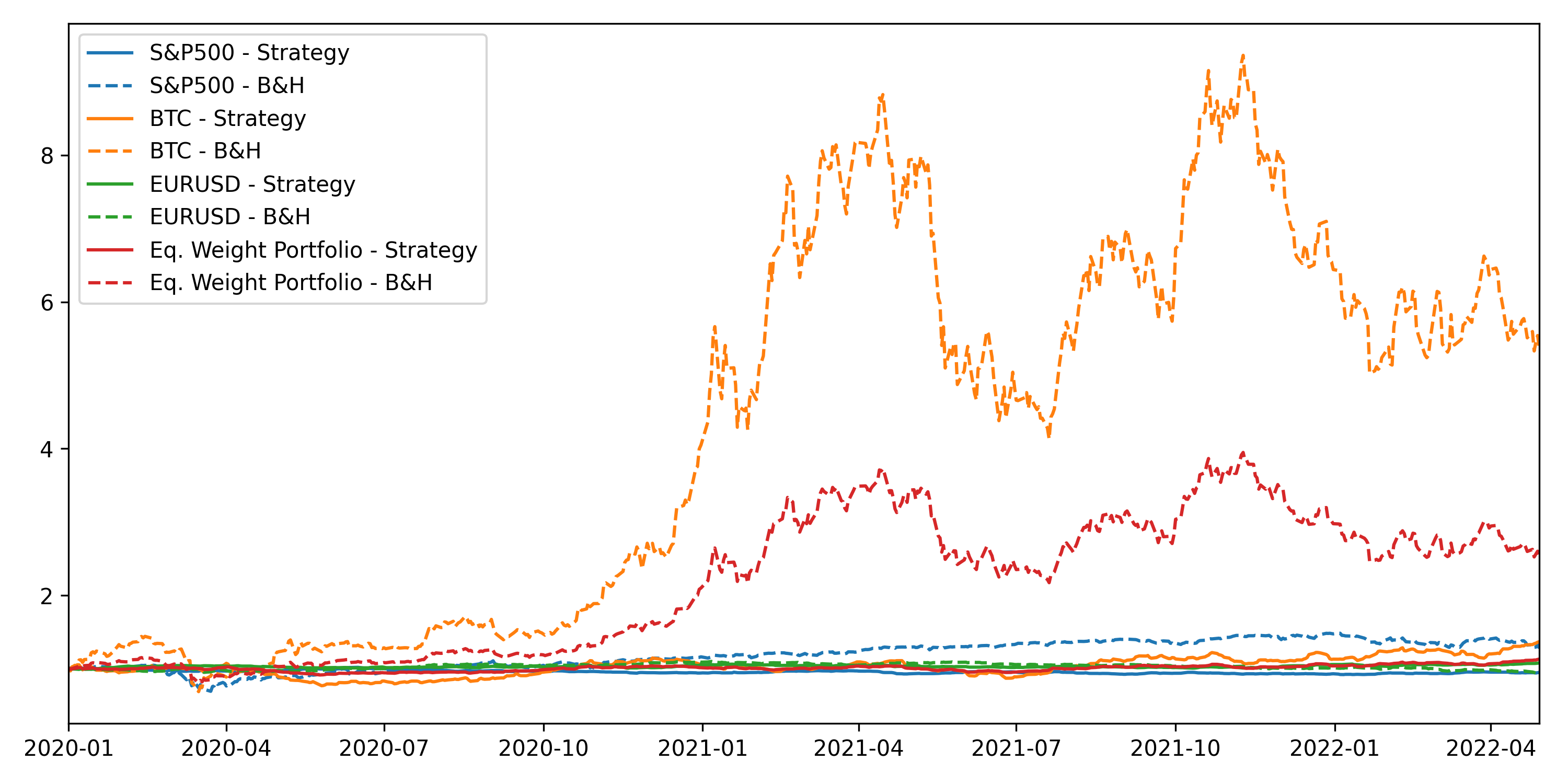}
    \label{fig:approach2_30min}
    \vspace{0.5cm} 
    \caption*{\footnotesize Source: Own Elaboration. Out-of-sample performance between 01.01.2020 and 30.04.2022, own backtesting implementation.}
\end{minipage}\hfill 
\begin{minipage}{0.45\textwidth}
    \centering
    \setstretch{1.1}
    \captionof{table}{Performance metrics: Approach 2 with 30-minute bar frequency (top) compared to buy and hold (bottom).}
    {\scriptsize 
    \begin{tabular}{lcccc}
    \toprule
    Approach 2 - 30min & SP500 & BTCUSD & EURUSD & Eq. Weighted \\
    \midrule
    Cumulative Return & -3.51\% & 32.51\% & 7.15\% & 12.05\% \\
    Annual Return & -1.47\% & 12.40\% & 2.91\% & 4.84\% \\
    Annualized Std & 4.89\% & 24.53\% & 4.85\% & 8.72\% \\
    Information Ratio & -0.30 & 0.51 & 0.60 & 0.55 \\
    Max Drawdown & 9.22\% & 30.44\% & 5.09\% & 11.04\% \\
    MLD & 455 days & 207 days & 251 days & 206 days \\
    IR** & -0.05 & 0.21 & 0.34 & 0.24 \\
    \midrule
    Buy And Hold & SP500 & BTCUSD & EURUSD & Eq. Weighted \\
    \midrule
    Cumulative Return & 26.83\% & 452.57\% & -5.58\% & 157.94\% \\
    Annual Return & 10.76\% & 108.57\% & -2.44\% & 50.30\% \\
    Annualized Std & 25.58\% & 69.97\% & 6.93\% & 45.38\% \\
    Information Ratio & 0.42 & 1.55 & -0.35 & 1.11 \\
    Max Drawdown & 38.25\% & 67.04\% & 15.78\% & 49.95 \% \\
    MLD & 125 days & 129 days & 331 days & 129 days \\
    IR** & 0.12 & 2.51 & -0.05 &1.12 \\
    \bottomrule
    \end{tabular}
    }
    \caption*{\footnotesize Source: Own Elaboration. Out-of-sample performance between 01.01.2020 and 30.04.2022, own backtesting implementation.} 
    \label{tab:approach2_30min}
\end{minipage}
\end{figure}

For approach 2 with 15-min bars (Figure 17, Table 9), Bitcoin (BTCUSD) achieved the highest cumulative return at 19.92\%, while the S\&P 500 (SP500) experienced a negative cumulative return of -3.08\%. In terms of annual return, Bitcoin again outperformed with 7.83\%, and the SP500 remained negative at -1.29\%. The annualized standard deviation was the highest for Bitcoin at 22.99\%, suggesting higher volatility, while EUR/USD exhibited the lowest volatility at 4.79\%.

Most crucially, when considering the $IR^{**}$ — which accounts for drawdown depth — the EUR/USD pair stands out with a robust score of 0.55, signifying a favorable risk-adjusted return after accounting for drawdowns. The equally-weighted portfolio had an Information Ratio** of 0.16, which, while lower than the individual EUR/USD strategy, still indicates a positive risk-adjusted performance. The negative Information Ratio** for the SP500 at -0.03 suggests an unfavorable return when adjusting for drawdown depth, highlighting the importance of this metric in evaluating strategy performance.

For Approach 2 with 30-minute bars (Figure 18, Table 10), Bitcoin leads the cumulative return at an impressive 32.51\%, outshining the SP500 which displayed a negative cumulative return of -3.51\%. The annual return follows a similar pattern, with Bitcoin achieving a substantial 12.40\% compared to the SP500’s -1.47\%, alongside EURUSD annual return of 2.91\%.

Most importantly, when factoring in the drawdown depth with the $IR^{**}$, Bitcoin and the equally-weighted portfolio exhibit positive figures of 0.21 and 0.24, indicating a better risk-adjusted return profile. The EUR/USD also maintains a strong $IR^{**}$ at 0.34. However, the SP500 continues to underperform on a risk-adjusted basis, reflected by its negative $IR^{**}$ of -0.05, further emphasizing the asset’s challenges within this particular approach and frequency setting.

\begin{figure}[t]
\begin{minipage}{0.45\textwidth}
    \centering
    \caption{Cumulative returns for Approach 3 with 5-minute bar frequency compared to buy and hold.}
    \includegraphics[width=\linewidth]{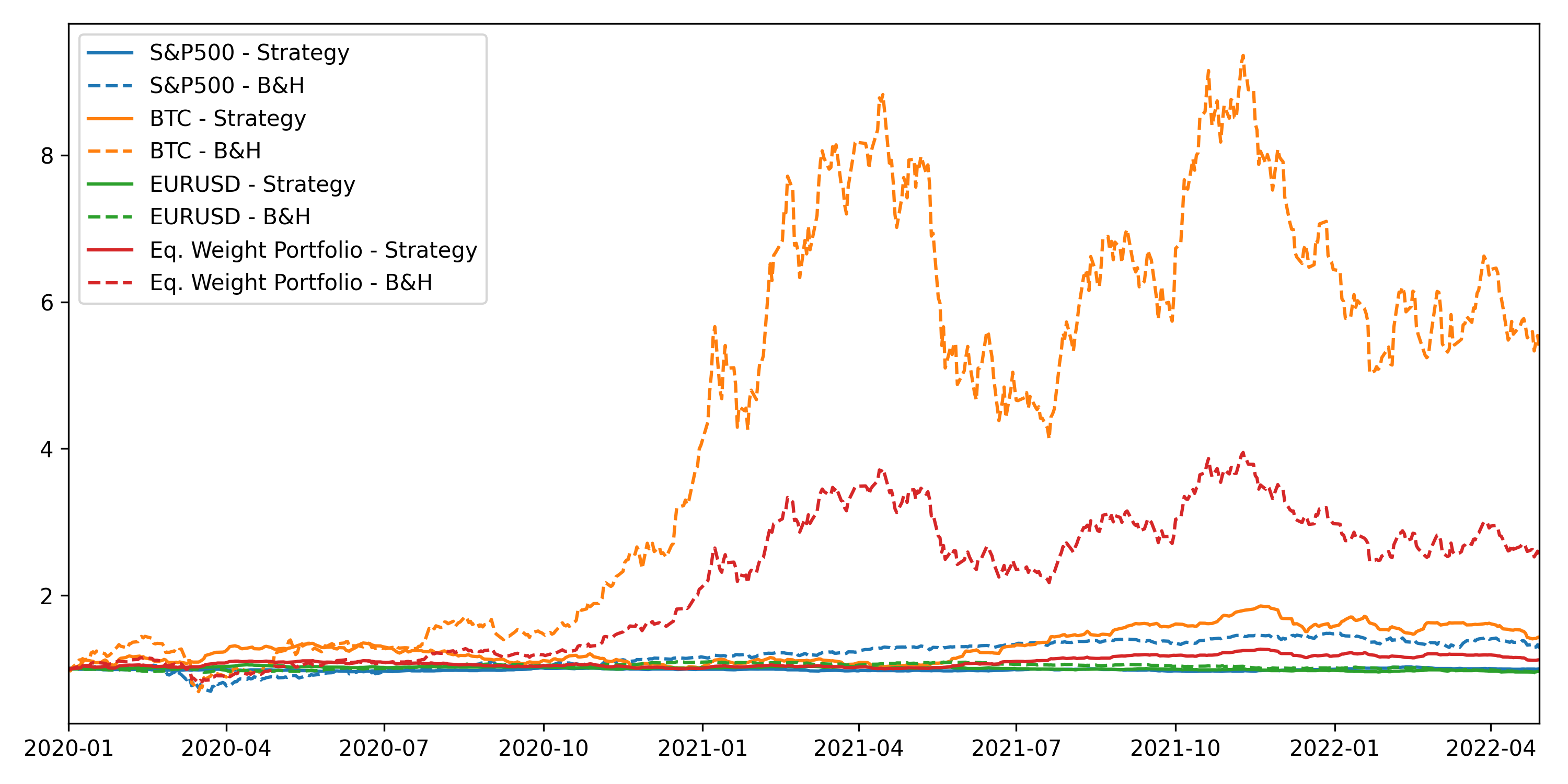}
    \label{fig:approach3_5min}
    \vspace{0.5cm} 
    \caption*{\footnotesize Source: Own Elaboration. Out-of-sample performance between 01.01.2020 and 30.04.2022, own backtesting implementation.}
\end{minipage}\hfill 
\begin{minipage}{0.45\textwidth}
    \centering
    \setstretch{1.1}
    \captionof{table}{Performance metrics: Approach 3 with 5-minute bar frequency (top) compared to buy and hold (bottom).}
    {\scriptsize 
    \begin{tabular}{lcccc}
    \toprule
    Approach 3 - 5min & SP500 & BTCUSD & EURUSD & Eq. Weighted \\
    \midrule
    Cumulative Return & 0.29\% & 41.29\% & -3.18\% & 12.80\% \\
    Annual Return & 0.12\% & 15.43\% & -1.33\% & 5.13\% \\
    Annualized Std & 4.68\% & 23.35\% & 4.66\% & 9.44\% \\
    Information Ratio & 0.03 & 0.66 & -0.29 & 0.54 \\
    Max Drawdown & 5.21\% & 45.75\% & 9.94\% & 15.74\% \\
    MLD & 350 days & 269 days & 520 days & 290 days \\
    IR** & 0.0 & 0.22 & -0.04 & 0.18 \\
    \midrule
    Buy And Hold & SP500 & BTCUSD & EURUSD & Eq. Weighted \\
    \midrule
    Cumulative Return & 26.83\% & 452.57\% & -5.58\% & 157.94\% \\
    Annual Return & 10.76\% & 108.57\% & -2.44\% & 50.30\% \\
    Annualized Std & 25.58\% & 69.97\% & 6.93\% & 45.38\% \\
    Information Ratio & 0.42 & 1.55 & -0.35 & 1.11 \\
    Max Drawdown & 38.25\% & 67.04\% & 15.78\% & 49.95 \% \\
    MLD & 125 days & 129 days & 331 days & 129 days \\
    IR** & 0.12 & 2.51 & -0.05 &1.12 \\
    \bottomrule
    \end{tabular}
    }
    \caption*{\footnotesize Source: Own Elaboration. Out-of-sample performance between 01.01.2020 and 30.04.2022, own backtesting implementation.} 
    \label{tab:approach3_5min}
\end{minipage}
\end{figure}

\subsection{Approach 3 - SAE-MLP Classifier}

Approach 3, employing a 5-minute bar frequency (Figure 19, Table 11), presents a varied performance landscape across different assets and the equally-weighted portfolio. The cumulative return for Bitcoin is notably high at 41.29\%, significantly outperforming the S\&P 500 index which shows a marginal gain of 0.29\%, and the EUR/USD currency pair that experienced a decline of -3.18\%. Bitcoin’s annual return stands at a compelling 15.43\%, while the SP500 and EUR/USD demonstrate a stark contrast with 0.12\% and -1.33\% respectively.

When considering the $IR^{**}$, Bitcoin maintains a positive score of 0.22, and the equally weighted portfolio also demonstrates resilience with a score of 0.18. The $IR^{**}$ for the SP500 breaks even at 0.0. The EUR/USD, however, has a slightly negative $IR^{**}$ of -0.04, reflecting a modestly unfavorable risk-adjusted return when taking drawdown depth into account.

\begin{figure}[t]
\begin{minipage}{0.45\textwidth}
    \centering
    \caption{Cumulative returns for Approach 3 with 15-minute bar frequency compared to buy and hold.}
    \includegraphics[width=\linewidth]{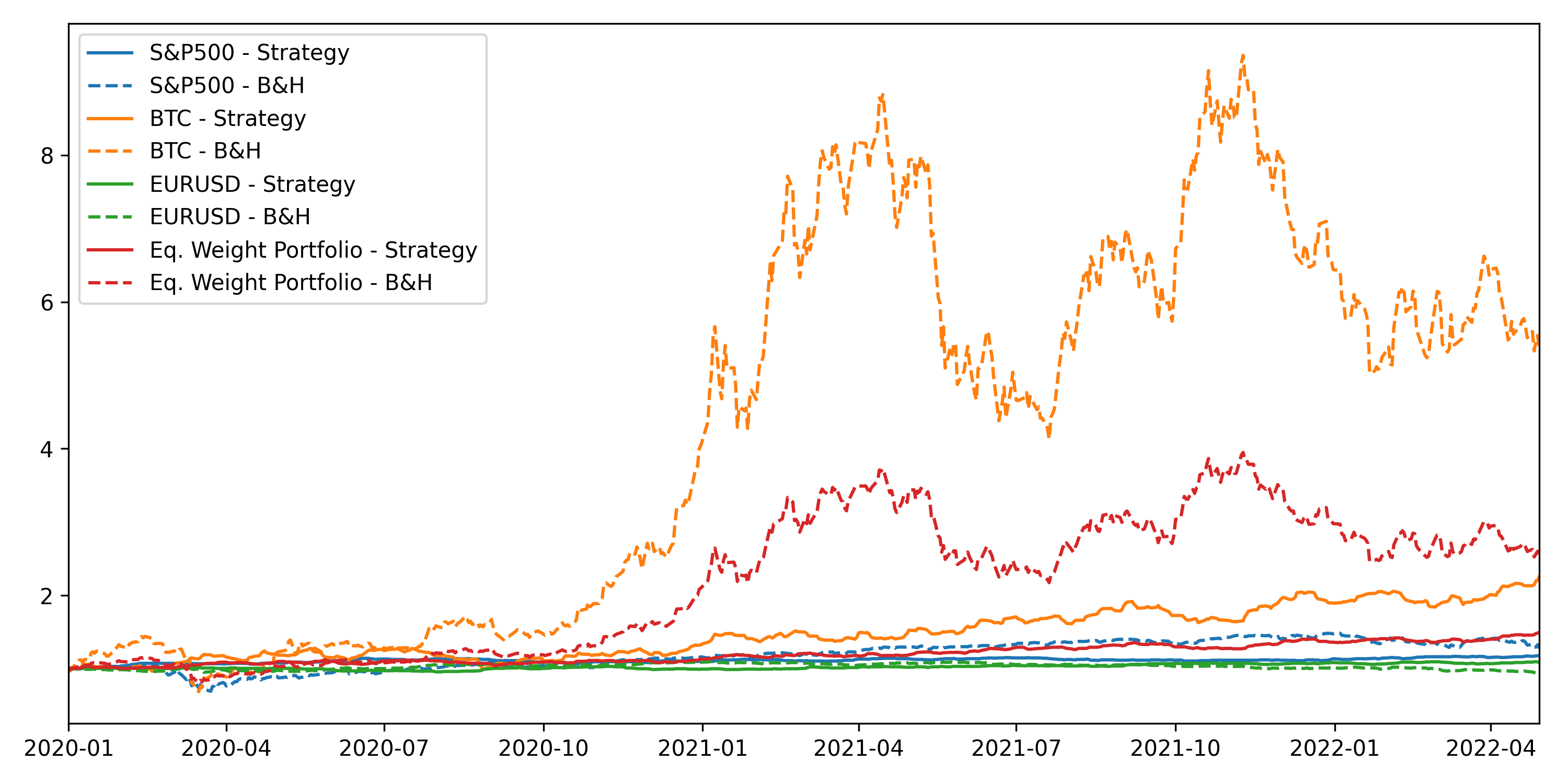}
    \label{fig:approach3_15min}
    \vspace{0.5cm} 
    \caption*{\footnotesize Source: Own Elaboration. Out-of-sample performance between 01.01.2020 and 30.04.2022, own backtesting implementation.}
\end{minipage}\hfill 
\begin{minipage}{0.45\textwidth}
    \centering
    \setstretch{1.1}
    \captionof{table}{Performance metrics: Approach 3 with 15-minute bar frequency (top) compared to buy and hold (bottom).}
    {\scriptsize 
    \begin{tabular}{lcccc}
    \toprule
    Approach 3 - 15min & SP500 & BTCUSD & EURUSD & Eq. Weighted \\
    \midrule
    Cumulative Return & 15.36\% & 115.28\% & 8.21\% & 46.28\% \\
    Annual Return & 6.11\% & 37.48\% & 3.33\% & 17.11\% \\
    Annualized Std & 5.09\% & 24.45\% & 4.81\% & 10.03\% \\
    Information Ratio & 1.20 & 1.53 & 0.69 & 1.71 \\
    Max Drawdown & 6.59\% & 28.65\% & 5.71\% & 9.07\% \\
    MLD & 199 days & 151 days & 150 days & 151 days \\
    IR** & 1.11 & 2.01 & 0.4 & 3.22 \\
    \midrule
    Buy And Hold & SP500 & BTCUSD & EURUSD & Eq. Weighted \\
    \midrule
    Cumulative Return & 26.83\% & 452.57\% & -5.58\% & 157.94\% \\
    Annual Return & 10.76\% & 108.57\% & -2.44\% & 50.30\% \\
    Annualized Std & 25.58\% & 69.97\% & 6.93\% & 45.38\% \\
    Information Ratio & 0.42 & 1.55 & -0.35 & 1.11 \\
    Max Drawdown & 38.25\% & 67.04\% & 15.78\% & 49.95 \% \\
    MLD & 125 days & 129 days & 331 days & 129 days \\
    IR** & 0.12 & 2.51 & -0.05 &1.12 \\
    \bottomrule
    \end{tabular}
    }
    \caption*{\footnotesize Source: Own Elaboration. Out-of-sample performance between 01.01.2020 and 30.04.2022, own backtesting implementation.} 
    \label{tab:approach3_15min}
\end{minipage}
\end{figure}

\begin{figure}[t]
\begin{minipage}{0.45\textwidth}
    \centering
    \caption{Cumulative returns for Approach 3 with 30-minute bar frequency compared to buy and hold.}
    \includegraphics[width=\linewidth]{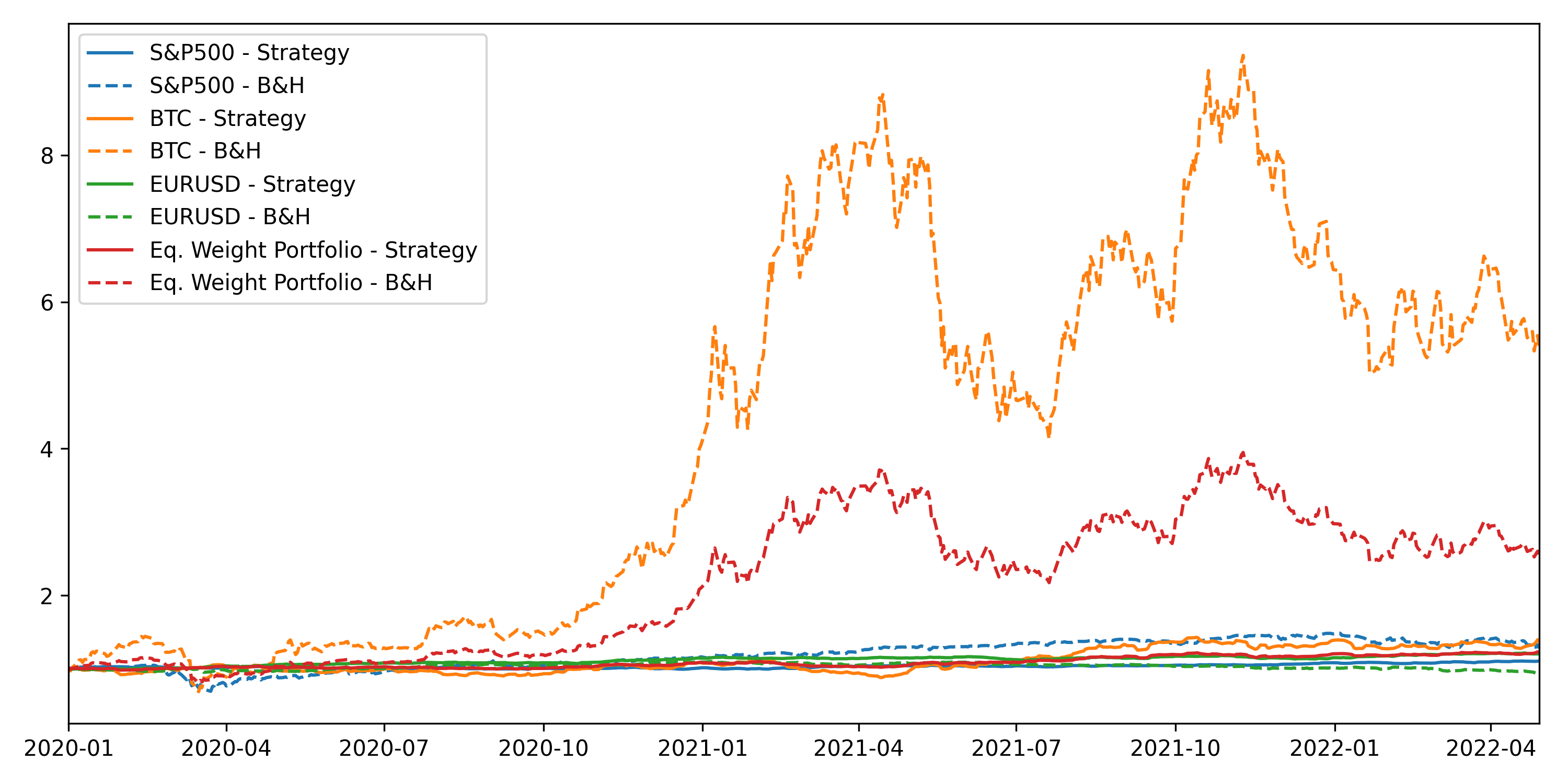}
    \label{fig:approach3_30min}
    \vspace{0.5cm} 
    \caption*{\footnotesize Source: Own Elaboration. Out-of-sample performance between 01.01.2020 and 30.04.2022, own backtesting implementation.}
\end{minipage}\hfill 
\begin{minipage}{0.45\textwidth}
    \centering
    \setstretch{1.1}
    \captionof{table}{Performance metrics: Approach 3 with 30-minute bar frequency (top) compared to buy and hold (bottom).}
    {\scriptsize 
    \begin{tabular}{lcccc}
    \toprule
    Approach 3 - 30min & SP500 & BTCUSD & EURUSD & Eq. Weighted \\
    \midrule
    Cumulative Return & 12.37\% & 39.88\% & 21.78\% & 24.68\% \\
    Annual Return & 4.96\% & 14.95\% & 8.53\% & 9.59\% \\
    Annualized Std & 4.78\% & 22.11\% & 4.91\% & 7.76\% \\
    Information Ratio & 1.04 & 0.68 & 1.74 & 1.24 \\
    Max Drawdown & 4.18\% & 29.18\% & 5.69\% & 8.48\% \\
    MLD & 240 days & 157 days & 91 days & 136 days \\
    IR** & 1.23 & 0.35 & 2.6 & 1.4 \\
    \midrule
    Buy And Hold & SP500 & BTCUSD & EURUSD & Eq. Weighted \\
    \midrule
    Cumulative Return & 26.83\% & 452.57\% & -5.58\% & 157.94\% \\
    Annual Return & 10.76\% & 108.57\% & -2.44\% & 50.30\% \\
    Annualized Std & 25.58\% & 69.97\% & 6.93\% & 45.38\% \\
    Information Ratio & 0.42 & 1.55 & -0.35 & 1.11 \\
    Max Drawdown & 38.25\% & 67.04\% & 15.78\% & 49.95 \% \\
    MLD & 125 days & 129 days & 331 days & 129 days \\
    IR** & 0.12 & 2.51 & -0.05 &1.12 \\
    \bottomrule
    \end{tabular}
    }
    \caption*{\footnotesize Source: Own Elaboration. Out-of-sample performance between 01.01.2020 and 30.04.2022, own backtesting implementation.} 
    \label{tab:approach3_30min}
\end{minipage}
\end{figure}

Approach 3 with a 15-minute bar frequency (Figure 20, Table 12) reveals significant improvements in performance metrics across the S\&P 500 (SP500), Bitcoin (BTCUSD), and the EUR/USD currency pair, as well as the equally-weighted portfolio of these assets. Bitcoin stands out with an exceptional cumulative return of 115.28\%, while the SP500 also reports a robust 15.36\% and the EUR/USD shows a moderate gain of 8.21\%. When annualized, Bitcoin's return remains impressive at 37.48\%, with the SP500 and EUR/USD yielding 6.11\% and 3.33\% respectively.

 Bitcoin's $IR^{**}$ at 2.01 and the equally-weighted portfolio's at 3.22 are high, suggesting an excellent return even when accounting for drawdown severity. The SP500 also maintains a strong $IR^{**}$ at 1.11. The EUR/USD exhibits a lower ratio of 0.4, indicating a less favorable return on risk when drawdowns are considered.

Approach 3 with a 30-minute bar frequency (Figure 21, Table 13) showcases a balanced performance across the assets. The cumulative return for Bitcoin is notable at 39.88\%, with the SP500 and EUR/USD also posting positive returns of 12.37\% and 21.78\%, respectively. The equally weighted portfolio benefits from the combined performance, yielding a cumulative return of 24.68\%.

Most importantly, when considering the $IR^{**}$, the EUR/USD achieves an exceptional score of 2.6, reflecting an outstanding risk-adjusted return considering drawdown depth. The equally weighted portfolio also performs impressively with an $IR^{**}$ of 1.4, further emphasizing the benefits of diversification. The SP500’s $IR^{**}$ stands at a solid 1.23, while Bitcoin lags with a score of 0.35, indicating that its higher returns are accompanied by proportionately higher risks and drawdowns. This data highlights the strength of the EUR/USD in this particular strategy and timeframe, outperforming even the robust returns of an equally weighted portfolio.

\subsection{Approach 4 - SAE-MLP + TBL Classification Forecasting}

\begin{figure}[t]
\begin{minipage}{0.45\textwidth}
    \centering
    \caption{Cumulative returns for Approach 4 with 5-minute bar frequency compared to buy and hold.}
    \includegraphics[width=\linewidth]{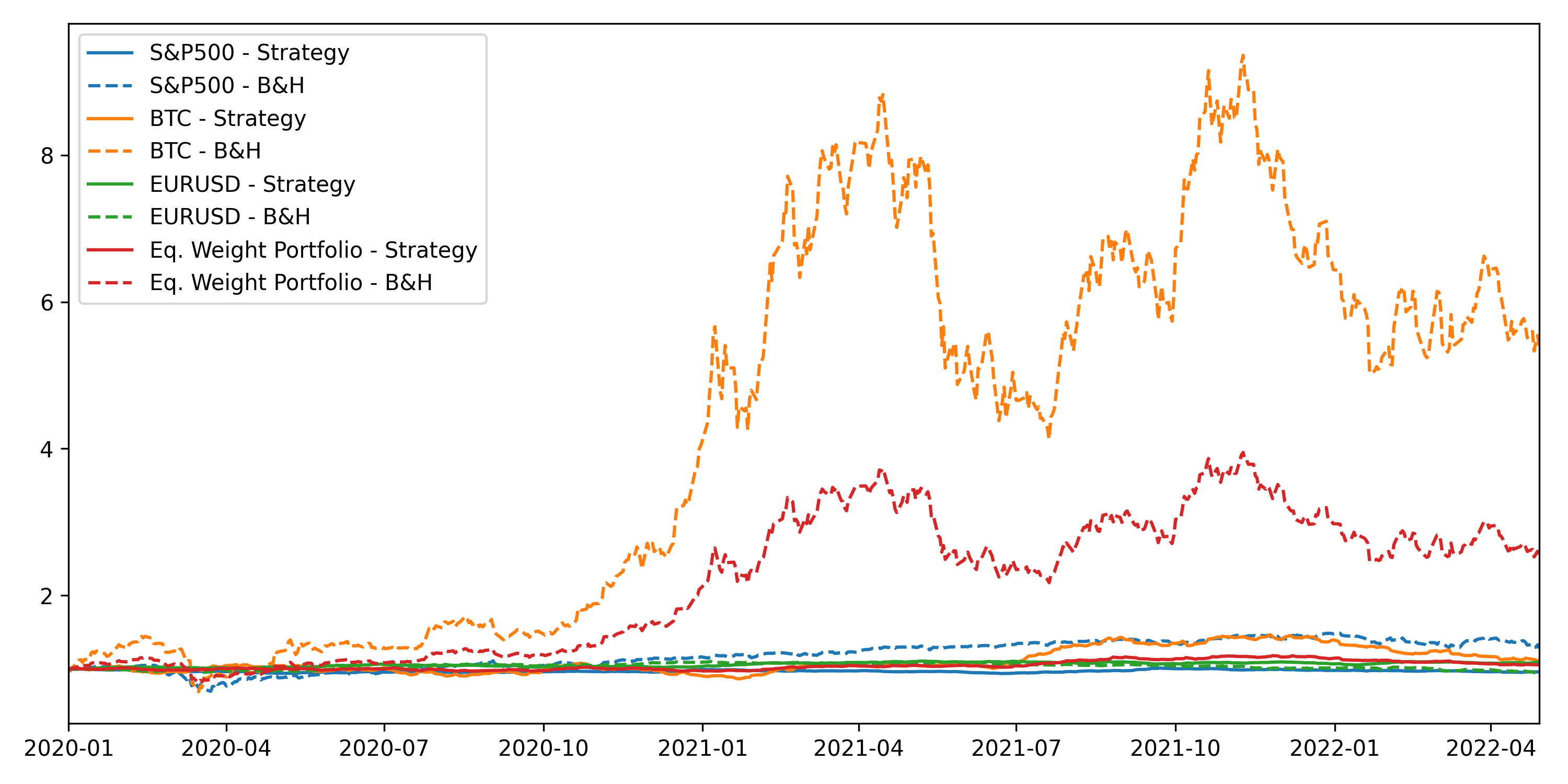}
    \label{fig:approach4_5min}
    \vspace{0.5cm} 
    \caption*{\footnotesize Source: Own Elaboration. Out-of-sample performance between 01.01.2020 and 30.04.2022, own backtesting implementation.}
\end{minipage}\hfill 
\begin{minipage}{0.45\textwidth}
    \centering
    \setstretch{1.1}
    \captionof{table}{Performance metrics: Approach 4 with 5-minute bar frequency (top) compared to buy and hold (bottom).}
    {\scriptsize 
    \begin{tabular}{lcccc}
    \toprule
    Approach 4 - 5min & SP500 & BTCUSD & EURUSD & Eq. Weighted \\
    \midrule
    Cumulative Return & -5.64\% & 11.78\% & 10.12\% & 5.42\% \\
    Annual Return & -2.38\% & 4.73\% & 4.08\% & 2.22\% \\
    Annualized Std & 4.59\% & 22.72\% & 4.84\% & 8.47\% \\
    Information Ratio & -0.52 & 0.21 & 0.84 & 0.26 \\
    Max Drawdown & 8.13\% & 39.53\% & 5.99\% & 14.37\% \\
    MLD & 462 days & 149 days & 248 days & 182 days \\
    IR** & -0.15 & 0.02 & 0.57 & 0.04 \\
    \midrule
    Buy And Hold & SP500 & BTCUSD & EURUSD & Eq. Weighted \\
    \midrule
    Cumulative Return & 26.83\% & 452.57\% & -5.58\% & 157.94\% \\
    Annual Return & 10.76\% & 108.57\% & -2.44\% & 50.30\% \\
    Annualized Std & 25.58\% & 69.97\% & 6.93\% & 45.38\% \\
    Information Ratio & 0.42 & 1.55 & -0.35 & 1.11 \\
    Max Drawdown & 38.25\% & 67.04\% & 15.78\% & 49.95 \% \\
    MLD & 125 days & 129 days & 331 days & 129 days \\
    IR** & 0.12 & 2.51 & -0.05 &1.12 \\
    \bottomrule
    \end{tabular}
    }
    \caption*{\footnotesize Source: Own Elaboration. Out-of-sample performance between 01.01.2020 and 30.04.2022, own backtesting implementation.} 
    \label{tab:approach4_5min}
\end{minipage}
\end{figure}

Approach 4 using a 5-minute bar frequency (Figure 22, Table 14) depicts a contrasting performance profile across the assets. The SP500 shows a cumulative return of -5.64\%, indicating a decline, whereas Bitcoin and EUR/USD both report positive cumulative returns of 11.78\% and 10.12\%, respectively. The equally weighted portfolio's cumulative return stands at 5.42\%, reflecting the mixed results of the underlying assets.

When evaluating the $IR^{**}$, the EUR/USD excels with a robust score of 0.57. The equally weighted portfolio and Bitcoin present marginal $IR^{**}$ scores of 0.04 and 0.02, respectively, reflecting limited risk-adjusted returns when accounting for drawdowns. The SP500's negative $IR^{**}$ of -0.15 further accentuates its underperformance in this approach. These results highlight the outperformance of the EUR/USD currency pair within Approach 4, especially when considering the risk associated with drawdowns.

\begin{figure}[t]
\begin{minipage}{0.45\textwidth}
    \centering
    \caption{Cumulative returns for Approach 4 with 15-minute bar frequency compared to buy and hold.}
    \includegraphics[width=\linewidth]{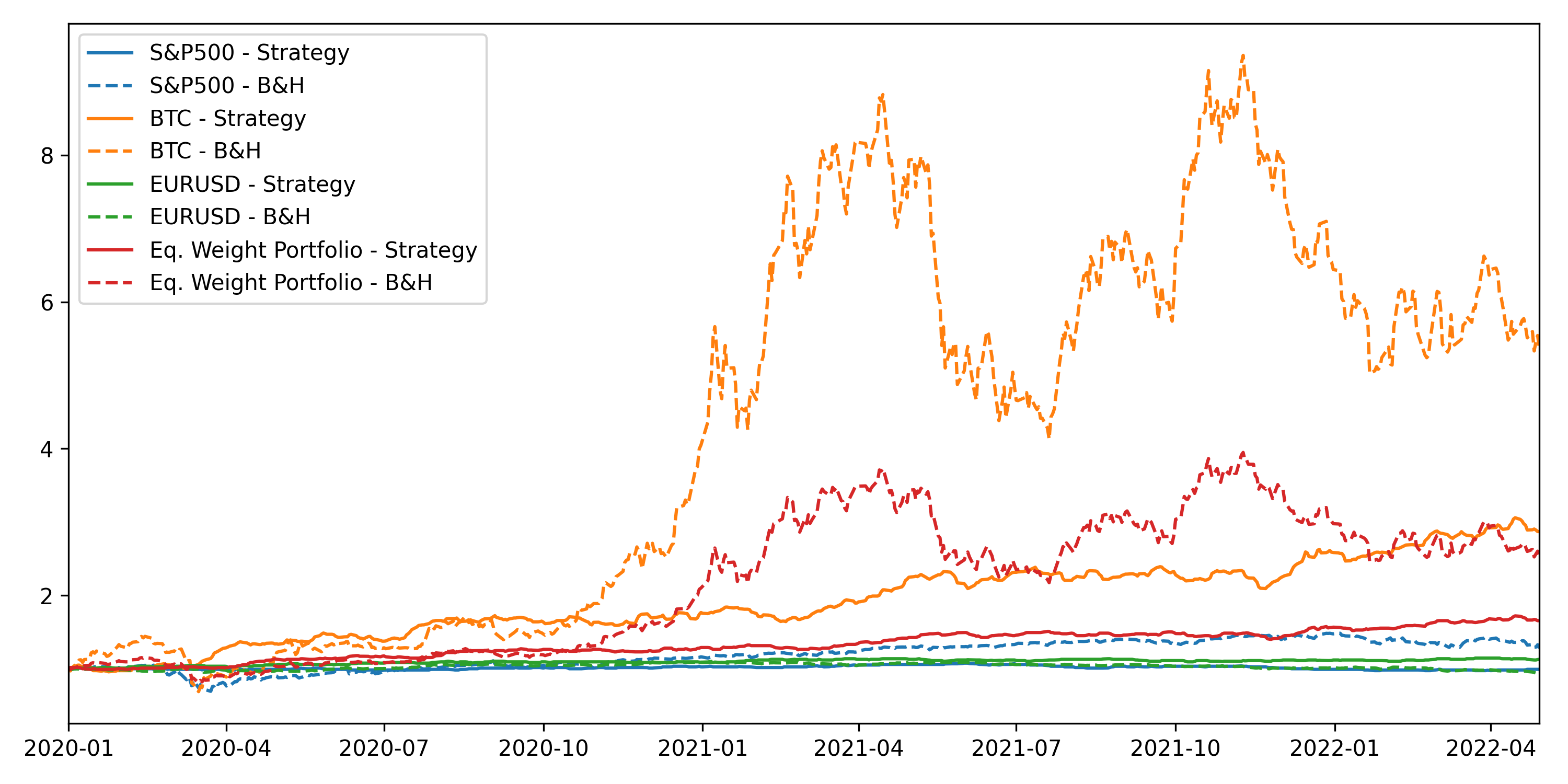}
    \label{fig:approach4_15min}
    \vspace{0.5cm} 
    \caption*{\footnotesize Source: Own Elaboration. Out-of-sample performance between 01.01.2020 and 30.04.2022, own backtesting implementation.}
\end{minipage}\hfill 
\begin{minipage}{0.45\textwidth}
    \centering
    \setstretch{1.1}
    \captionof{table}{Performance metrics: Approach 4 with 15-minute bar frequency (top) compared to buy and hold (bottom).}
    {\scriptsize 
    \begin{tabular}{lcccc}
    \toprule
    Approach 4 - 15min & SP500 & BTCUSD & EURUSD & Eq. Weighted \\
    \midrule
    Cumulative Return & -0.90\% & 180.77\% & 13.80\% & 64.55\% \\
    Annual Return & -0.38\% & 53.51\% & 5.51\% & 22.97\% \\
    Annualized Std & 4.60\% & 23.56\% & 4.70\% & 11.13\% \\
    Information Ratio & -0.08 & 2.27 & 1.17 & 2.06 \\
    Max Drawdown & 10.17\% & 30.09\% & 4.01\% & 10.79\% \\
    MLD & 222 days & 58 days & 220 days & 103 days \\
    IR** & -0.0 & 4.04 & 1.61 & 4.39 \\
    \midrule
    Buy And Hold & SP500 & BTCUSD & EURUSD & Eq. Weighted \\
    \midrule
    Cumulative Return & 26.83\% & 452.57\% & -5.58\% & 157.94\% \\
    Annual Return & 10.76\% & 108.57\% & -2.44\% & 50.30\% \\
    Annualized Std & 25.58\% & 69.97\% & 6.93\% & 45.38\% \\
    Information Ratio & 0.42 & 1.55 & -0.35 & 1.11 \\
    Max Drawdown & 38.25\% & 67.04\% & 15.78\% & 49.95 \% \\
    MLD & 125 days & 129 days & 331 days & 129 days \\
    IR** & 0.12 & 2.51 & -0.05 &1.12 \\
    \bottomrule
    \end{tabular}
    }
    \caption*{\footnotesize Source: Own Elaboration. Out-of-sample performance between 01.01.2020 and 30.04.2022, own backtesting implementation.} 
    \label{tab:approach4_15min}
\end{minipage}
\end{figure}

\begin{figure}[t]
\begin{minipage}{0.45\textwidth}
    \centering
    \caption{Cumulative returns for Approach 4 with 30-minute bar frequency compared to buy and hold.}
    \includegraphics[width=\linewidth]{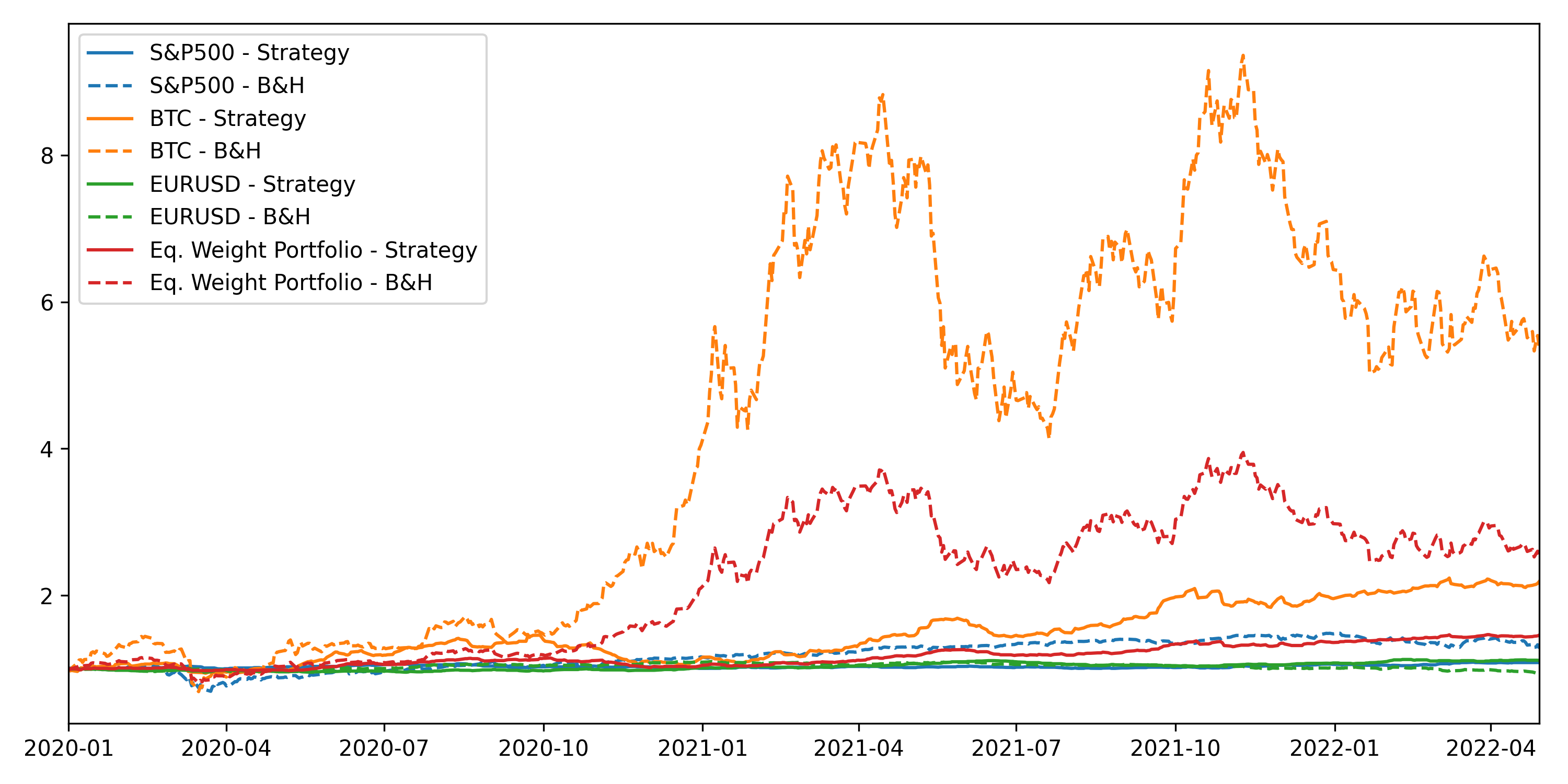}
    \label{fig:approach4_30min}
    \vspace{0.5cm} 
    \caption*{\footnotesize Source: Own Elaboration. Out-of-sample performance between 01.01.2020 and 30.04.2022, own backtesting implementation.}
\end{minipage}\hfill 
\begin{minipage}{0.45\textwidth}
    \centering
    \setstretch{1.1}
    \captionof{table}{Performance metrics: Approach 4 with 30-minute bar frequency (top) compared to buy and hold (bottom).}
    {\scriptsize 
    \begin{tabular}{lcccc}
    \toprule
    Approach 4 - 30min & SP500 & BTCUSD & EURUSD & Eq. Weighted \\
    \midrule
    Cumulative Return & 8.49\% & 116.30\% & 10.82\% & 45.20\% \\
    Annual Return & 3.44\% & 37.75\% & 4.36\% & 16.75\% \\
    Annualized Std & 4.84\% & 24.42\% & 4.92\% & 10.44\% \\
    Information Ratio & 0.71 & 1.55 & 0.89 & 1.60 \\
    Max Drawdown & 6.41\% & 39.74\% & 7.90\% & 12.54\% \\
    MLD & 383 days & 140 days & 255 days & 130 days \\
    IR** & 0.38 & 1.47 & 0.49 & 2.14 \\
    \midrule
    Buy And Hold & SP500 & BTCUSD & EURUSD & Eq. Weighted \\
    \midrule
    Cumulative Return & 26.83\% & 452.57\% & -5.58\% & 157.94\% \\
    Annual Return & 10.76\% & 108.57\% & -2.44\% & 50.30\% \\
    Annualized Std & 25.58\% & 69.97\% & 6.93\% & 45.38\% \\
    Information Ratio & 0.42 & 1.55 & -0.35 & 1.11 \\
    Max Drawdown & 38.25\% & 67.04\% & 15.78\% & 49.95 \% \\
    MLD & 125 days & 129 days & 331 days & 129 days \\
    IR** & 0.12 & 2.51 & -0.05 &1.12 \\
    \bottomrule
    \end{tabular}
    }
    \caption*{\footnotesize Source: Own Elaboration. Out-of-sample performance between 01.01.2020 and 30.04.2022, own backtesting implementation.} 
    \label{tab:approach4_30min}
\end{minipage}
\end{figure}

Approach 4 with a 15-minute bar frequency (Figure 23, Table 15) presents a remarkable divergence in the performance of the SP500, Bitcoin (BTCUSD), EUR/USD, and an equally-weighted portfolio. Bitcoin commands the stage with an extraordinary cumulative return of 180.77\%, dwarfing the SP500's slight decline of -0.90\% and surpassing the solid performance of EUR/USD at 13.80\%. The equally weighted portfolio benefits from Bitcoin's stellar performance, achieving a cumulative return of 64.55\%.

When considering the Information Ratio with two asterisks, which adjusts for drawdown depth, Bitcoin exhibits a score of 4.04, and the equally-weighted portfolio surpasses this with a ratio of 4.39, both indicating high risk-adjusted returns. The EUR/USD also shows a strong $IR^{**}$ of 1.61. In contrast, the SP500 breaks even with a ratio of -0.0, suggesting no excess return after adjusting for drawdowns. These results highlight the potential benefits of combining high-return assets with diversification strategies.

Approach 4 with a 30-minute bar frequency (Figure 24, Table 16) demonstrates a varied performance among the SP500, Bitcoin (BTCUSD), EUR/USD, and an equally weighted portfolio. Bitcoin delivers a striking cumulative return of 116.30\%, significantly outshining the SP500's return of 8.49\% and EUR/USD's return of 10.82\%. The equally weighted portfolio presents an impressive cumulative return of 45.20\%, reflecting the strong performance of Bitcoin.

When adjusted for drawdown depth with the $IR^{**}$, Bitcoin's risk-adjusted performance is still strong at 1.47, but it is the equally-weighted portfolio that stands out with a ratio of 2.14, showcasing excellent risk-adjusted returns. The SP500 and EUR/USD have $IR^{**}$ of 0.38 and 0.49, respectively, which are positive but less remarkable compared to the portfolio's.

\subsection{Summary}

Our research has demonstrated that the application of Supervised Autoencoder de-noising in combination with Triple Barrier labeling significantly improves machine learning performance in the context of financial time series analysis. Figure 25 demonstrates that Supervised Autoencoder allows for effective de-noising of the data, essentially enhancing the signal-to-noise ratio, and thus leading to better model performance. Table 17 presents p-values for two statistical tests that evaluate whether the strategy produced statistically better performance than the buy-and-hold strategies.

\begin{figure}
  \centering
  \caption{Comparison of information ratios across approaches and bar lengths}
      \makebox[\columnwidth][c]{\includegraphics[width=0.6\columnwidth]{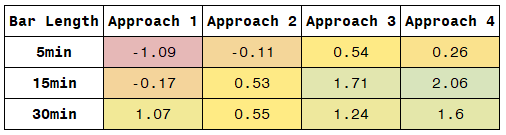}}%
      \caption*{\footnotesize Source: Own Elaboration, table produced in Microsoft Excel.}
\end{figure}

The Diebold-Mariano test (Diebold, Mariano, 2002), is specifically designed to evaluate and compare the relative forecasting accuracy of two competing predictive models. This test calculates the differences in the predictive errors of the two models and assesses whether these differences are statistically significant. The test assumes that these differences are normally distributed, especially under large sample sizes, in accordance with the central limit theorem.

On the other hand, we also employed a probabilistic t-test to compare the Information ratios of the two investment strategies. This test utilizes a simplified formula for the standard error of the Sharpe ratio, expressed as $SE = \frac{\sigma}{\sqrt{n}}$, where $n$ is the sample size of the returns, and  $\sigma$  is the standard deviation of the differences in returns between the two investment strategies. The complete test statistic can be expressed as:

\begin{equation}
    t = \frac{IR_1 - IR_2}{\frac{\sigma}{\sqrt{n}}}
\end{equation}

Like the Diebold-Mariano test, the probabilistic t-test assumes that the differences in returns follow a normal distribution. Both tests, the DM test, and the probabilistic t-test, are essential to assess the effectiveness of our proposed approach. We set a critical value $\alpha = 0.01$ for these tests, indicating a strict criterion for statistical significance. 

The table indicates that the most tests where the strategy significantly outperformed buy and hold were trading EUR/USD and SP500, whereas it was particularly difficult to achieve similar results for BTC/USD, as only one test (Approach 4, 15min) significantly outperformed said cryptocurrency. In terms of these statistical tests, this approach was also the most successful one, rejecting the null hypotheses in 5 out of 8 tests.

What is also important to note, is that for equally weighted portfolio strategies, all approaches that succeeded in rejecting the null hypotheses involved the use of SAE or triple barrier labeling. Based on these results, we view that triple barrier labeling provides a robust mechanism to label data based on predetermined profit-taking and stop-loss levels, capturing more realistic and complex market dynamics compared to traditional methods. This amalgamation of techniques improves model accuracy and predictive power, as well as handles financial market volatility and noise better.

\begin{table}
  \centering
  \caption{P-values for statistical tests for comparing the performance of strategies over buy \& hold.}
      \makebox[\columnwidth][c]{\includegraphics[width=0.6\columnwidth]{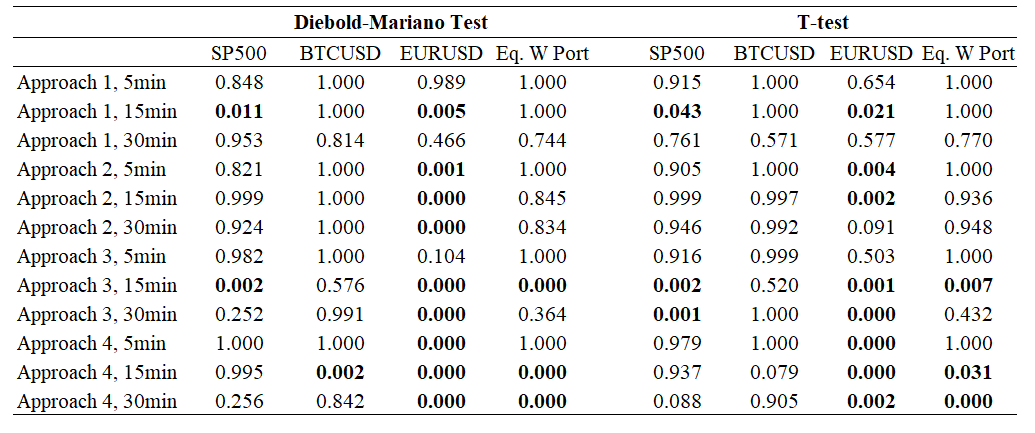}}%
      \caption*{\footnotesize Source: Own Elaboration, tests conducted with stattests package. DM $H_0$: Classifier accuracy is not greater than "always-long" accuracy. T-test $H_0$: The information Ratio of the strategy is not greater than the buy-and-hold information ratio. The bold values indicate approaches and assets where a given p-value exceeded the critical value, rejecting the null hypothesis.}
\end{table}

\section{Sensitivity Analysis}

The sensitivity analysis presented here concerns the approach 4, 15-minute bars. We view that the sensitivity analysis of this approach demonstrates particularly well the benefits and challenges of the SAE-MLP + TBL method. 

\begin{figure}
\caption{Sensitivity analysis of triple-barrier labeling parameters in Approach 4, 15min bars.}
  \centering
      \makebox[\columnwidth][c]{\includegraphics[width=0.8\columnwidth]{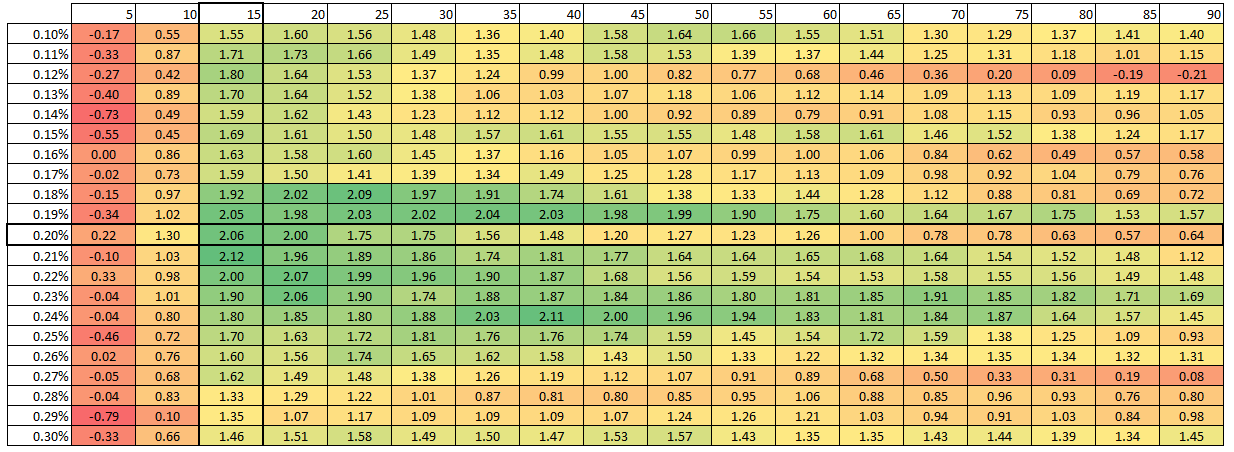}}%
            \caption*{\footnotesize Source: Own Elaboration. X-Axis: window length in minutes. Y-Axis: window height.}
\end{figure}

For this approach, the information ratio presented in Figure 26 increased as the stop-loss/take-profit levels were expanded up to around 0.18-0.21\%, beyond which we note a decline to 1.3-1.8 levels. This diminishing return could be attributed to overexposure to market volatility at higher stop-loss/take-profit levels, thus negatively affecting the risk-adjusted return. Furthermore, it is also evident that augmenting the trade duration up to an optimal point of approximately 15-20 minutes enhances the strategy performance. However, beyond this point, a further increase in trade duration does not uniformly enhance the performance, signifying increased uncertainty or noise over extended trading horizons. The observed variation in the information ratio underlines the strategy's sensitivity to these parameters, emphasizing the necessity for their careful calibration.

\begin{figure}
\caption{Sensitivity analysis of SAE parameters.}
  \centering
      \makebox[\columnwidth][c]{\includegraphics[width=0.6\columnwidth]{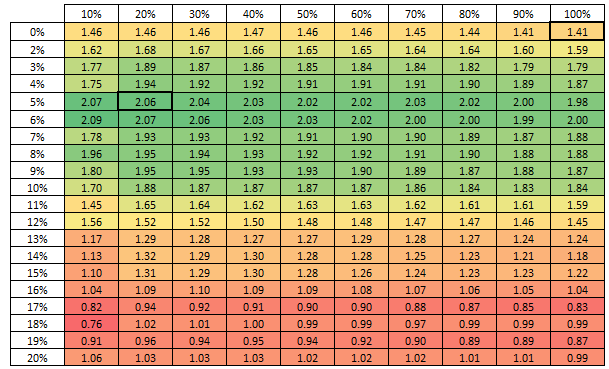}}%
            \caption*{\footnotesize Source: Own Elaboration, Y-axis: Noise rate as a fraction of annualized volatility. X-axis: bottleneck size as a percentage of original features count.}
\end{figure}

From Figure 27, it can be observed that the Information Ratio appears to be stable around the point where 5\% Gaussian noise is added, whereas the size of the autoencoder's bottleneck has much less impact on the overall information ratio. This suggests that a moderate amount of data augmentation, combined with a considerable reduction of feature size through the autoencoder, optimally improves the strategy's risk-adjusted return.

As the percentage of Gaussian noise added to features increases beyond 5\%, the Information Ratio generally decreases, indicating a potential overfitting problem. Too much noise may be causing the model to learn from this 'noise', which doesn't necessarily represent any true underlying pattern in the market data, leading to degradation of out-of-sample performance.

Similarly, if the size of the autoencoder's bottleneck is expanded beyond 40\% of the original number of features, the Information Ratio also shows a downward trend, albeit much less steep than in the case of the noise parameter. This could be due to the fact that too many features may retain more noise than signal, resulting in a less robust model.

\begin{figure}
\caption{Sensitivity analysis of SAE hidden layer count.}
  \centering
      \makebox[\columnwidth][c]{\includegraphics[width=0.5\columnwidth]{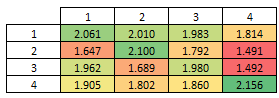}}%
            \caption*{\footnotesize Source: Own Elaboration. Y Axis - encoder hidden layer count. X Axis - decoder hidden layer count.}
\end{figure}

When it comes to the count of hidden layers in the autoencoder, figure 28 shows the optimal configuration is not uniformly increasing with the number of layers. The performance does not consistently improve with additional layers, indicating that an increase in model complexity beyond a certain point does not necessarily translate to better performance and may lead to overfitting or other inefficiencies. This observation aligns with the principle of parsimony in model selection, where the simplest model that adequately explains the data is preferred.

\begin{figure}
\caption{Sensitivity analysis of batch size and learning rate.}
  \centering
      \makebox[\columnwidth][c]{\includegraphics[width=0.8\columnwidth]{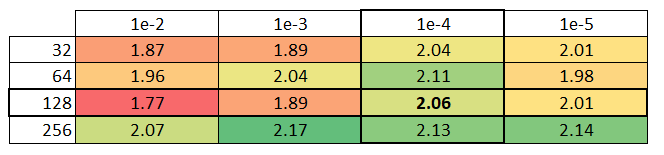}}%
            \caption*{\footnotesize Source: Own Elaboration. Y-Axis - batch size. X-Axis - learning rate.}
\end{figure}

Figure 29 presents the sensitivity analysis of batch sizes and learning rates, A notable pattern is that larger batch sizes tend to achieve higher information ratios at lower learning rates, with the batch size of 256 at a learning rate of 1e-3 standing out as particularly effective. Likewise, the highest learning rates seem to produce on average worse results. This could suggest that smaller learning rates may help in avoiding oscillations around the global minimum, a phenomenon that can occur at higher learning rates, leading to a potential decrease in the information ratio.

In conclusion, the results imply that there is an optimal level of complexity for the autoencoder architecture that maximizes the Sharpe ratio. Consequently, careful consideration must be given to the selection of the number of hidden layers when designing such a strategy to achieve a balance between model expressiveness and performance.

\section*{Conclusions and Further Research}
\addcontentsline{toc}{section}{Conclusions and Further Research}

This research set out to explore the potential improvements in strategy performance yielded by the application of supervised autoencoders in the context of financial time series. We further incorporated noise augmentation and triple barrier labeling to understand the interplay of these factors in determining the strategy's risk-adjusted return, quantified by the Sharpe Ratio and the Information Ratio.

Through rigorous testing and sensitivity analysis, we demonstrated that employing supervised autoencoders significantly enhanced the performance metrics of the strategy. Our analysis suggested an optimal balance between the level of Gaussian noise added to the features and the size of the autoencoder's bottleneck.

However, an increase in the level of noise and bottleneck size beyond certain thresholds led to a decrease in performance, presumably due to overfitting and the inclusion of more noise than signal, respectively. These findings underline the need for careful calibration of these parameters to ensure the most effective utilization of supervised autoencoders in this context.

\begin{itemize}
    \item RQ1. Does data augmentation and denoising via autoencoders improve the performance of a strategy? - Findings presented in the results section (Figure 26) suggest that data augmentation using Gaussian noise and denoising via autoencoders significantly improved the performance of the strategy. This was evident since Approach 3 outperformed Approaches 1 and 2 across all bar lengths in terms of the Information Ratio of an equally weighted portfolio, under the optimal levels of noise and autoencoder bottleneck size. However, caution is necessary as the relationship between noise level, bottleneck size, and performance was not linear, indicating the need for careful calibration of these parameters.
    \item RQ2. Does triple barrier labeling improve classifier performance over simple direction classification? - Triple barrier labeling generally outperformed simple labeling due to its ability to handle market noise better (Figure 25), as well as the symmetry of rewards which makes for better optimization metrics, which is demonstrated by approach 4 outperforming approaches 1-3 in 15-minute and 30-minute bars. We note that triple barrier labeling may not be the best choice for high-frequency trading since the 5-minute bar performance was worse than for approach 3.
    \item RQ3. Does hyperparameter tuning help achieve better performance of the investment strategy? - Hyperparameter tuning was shown to be crucial in achieving superior performance in our strategy (Figures 26, 27, 28). The optimal performance was observed under a specific combination of noise level and autoencoder bottleneck size, emphasizing the importance of hyperparameter optimization in the application of machine learning techniques to trading strategies.
\end{itemize}

Despite the promising results, the study is subject to certain limitations. Firstly, the findings are based on historical data, and as such, their predictive power in relation to future performance should be considered with caution, given the volatile and evolving nature of financial markets. Secondly, the research does not take into account slippage, which could potentially diminish the net returns of the strategy, especially if dealing in illiquid markets or with large capital. The study also assumes that stop-losses and take profits from triple barrier labeling are executed immediately and perfectly, which is not always the case in the markets.

This paper introduces several new approaches to algorithmic trading. First, it appears to be the first study to apply our specific model architecture in the field of algorithmic trading. This approach differs notably from the traditional models typically seen in this area. Second, while the concept of triple barrier labeling has been previously discussed, our research goes a step further by developing a specialized optimization metric designed explicitly for use with triple barrier labeling. These contributions are significant steps forward in integrating advanced machine-learning techniques into financial trading strategies.

The empirical evidence indicating that algorithmic models can outperform traditional buy-and-hold strategies suggests a necessity for their adoption in asset management to enhance market efficiency and potential investment returns. Consequently, we view it as vital for regulators to craft policies that facilitate the ethical integration of these models, ensuring market fairness and stability while mitigating systemic risks. Institutional investors and fund managers are encouraged to embrace these advanced strategies, necessitating investments in technology and skilled personnel to maintain competitiveness and uphold their fiduciary responsibilities. This shift towards algorithmic trading is not only a reflection of the potential for improved financial performance but also a movement toward the inevitable modernization of financial market practices.

In terms of further research, we recommend investigating other types of noise and their impacts on the strategy's performance. Additionally, the integration of slippage into the model would provide a more realistic picture of the strategy's net returns. Exploring different architectures for the autoencoder or the integration of other deep learning techniques may also yield interesting insights. The impact of these methods on other types of financial time series data, beyond the one used in this study, would also be a fruitful avenue for future research.

The findings from this study provide a compelling case for the continued exploration of machine learning techniques in financial time series analysis and trading strategy development. As our understanding of these tools deepens, we move closer to unlocking their full potential in predictive modeling and decision-making within the complex landscape of financial markets.

\end{document}